\documentclass[aps,pra,reprint,showpacs,superscriptaddress,floatfix]{revtex4-2}
\usepackage[T1]{fontenc}
\usepackage[utf8]{inputenc}
\usepackage{lmodern}
\usepackage{amsmath,amssymb,amsthm,physics}
\usepackage{graphicx,epstopdf}
\usepackage[dvipsnames]{xcolor}
\usepackage{float}
\usepackage{textgreek}
\usepackage{ulem}
\usepackage{orcidlink}
\usepackage{appendix}
\allowdisplaybreaks

\hypersetup{
    colorlinks=true,
    linkcolor=blue,
    citecolor=blue,
    filecolor=blue,
    urlcolor=blue,
    breaklinks=true
}
\usepackage{silence}
\begin{document}

\title{ Accelerating Bertotti-Robinson Black Holes in a Uniform Magnetic Field }

\author{Ahmad Al-Badawi\orcidlink{0000-0002-3127-3453}}
\email{ahmadbadawi@ahu.edu.jo}
\affiliation{Department of Physics, Al-Hussein Bin Talal University, 71111, Ma'an, Jordan}

\author{Faizuddin Ahmed\orcidlink{0000-0003-2196-9622}}
\email{faizuddinahmed15@gmail.com}
\affiliation{Department of Physics, The Assam Royal Global University, Guwahati-781035, Assam, India}

\author{Edilberto O. Silva\orcidlink{0000-0002-0297-5747}}
\email{edilberto.silva@ufma.br}
\affiliation{Programa de P\'os-Gradua\c c\~ao em F\'{\i}sica \& Coordena\c c\~ao do Curso de F\'{\i}sica -- Bacharelado, Universidade Federal do Maranh\~{a}o, 65085-580 S\~{a}o Lu\'{\i}s, Maranh\~{a}o, Brazil}

\date{\today}

\begin{abstract}
We study the Hawking temperature, geodesic motion, and observable signatures of the accelerating Bertotti-Robinson (BR) spacetime, a vacuum black-hole solution deformed by a uniform magnetic field $B$ and an acceleration parameter $\alpha$. In the timelike sector, we derive the effective potential for massive particles, determine the specific energy and angular momentum for equatorial circular orbits, and determine how $(B,\alpha)$ shifts the ISCO; we also illustrate representative trajectories of massive particles. We then compute the radial and latitudinal epicyclic frequencies for small perturbations about circular orbits, quantifying how the magnetic field and acceleration modify local radial and vertical stability. In the null sector, we derive the photon effective potential and obtain analytical expressions for the photon-sphere radius, critical impact parameter, and shadow radius, complemented by photon trajectories, the effective radial force, and the Lyapunov exponent controlling the instability of circular null orbits; we also provide parameter-space maps for the photon sphere and shadow. Finally, we obtain the energy emission rate emitted from the black hole, showing how the acceleration parameter and the magnetic field affects this.\\

\end{abstract}

\maketitle

\tableofcontents

\section{Introduction}\label{sec1}

The discovery of the Bertotti-Robinson (BR) geometry as an exact solution to the Einstein-Maxwell equations established one of the earliest and most important examples of a spacetime with a constant electromagnetic field \cite{Bertotti1959, Robinson1959}. The BR solution, characterized by an $\mathrm{AdS}_2 \times S^2$ topology with uniform electric or magnetic field lines threading the two-sphere, has since found widespread applications in theoretical physics, ranging from the near-horizon geometry of extremal Reissner-Nordstr\"{o}m black holes to the study of string compactifications and holography.

Black holes immersed in external magnetic fields have been extensively studied in the literature. In particular, the magnetized Schwarzschild solution was analyzed in \cite{Ernst1976}, while the magnetized Kerr black hole was investigated in \cite{ErnstWild1976}. The properties of a cylindrical electromagnetic universe were previously examined in \cite{Melvin1965}. Exact solutions describing spacetimes with a uniform magnetic field and a magnetic dipole were obtained in \cite{Bonnor1954}. Furthermore, the electromagnetic field generated when a stationary, axisymmetric black hole is placed in an initially uniform magnetic field aligned with its symmetry axis was derived in \cite{Wald1974}.

A natural extension of this geometry, namely, the embedding of a Schwarzschild black hole into a BR background, was presented earlier in \cite{Halilsoy1993}. In a similar spirit, the Reissner-Nordström black hole can be consistently coupled to an external, aligned uniform electromagnetic field of the BR spacetime under appropriate equilibrium conditions \cite{Halilsoy1998}. Subsequently, a metric describing a Schwarzschild black hole interacting with an external, stationary electromagnetic BR field was derived in \cite{Halilsoy1995}. This solution generalizes the earlier construction of \cite{Halilsoy1993}, which aimed to unify the Schwarzschild and BR geometries within a common framework. Notably, certain limiting cases of the metric in \cite{Halilsoy1995} correspond to a stationary electromagnetic universe in which conformal curvature emerges because of rotation \cite{Halilsoy2004}.

Recently, Podolsk\'{y} et al. \cite{Hryhorii2025,Podolsky2025} introduced an exact solution describing a Kerr black hole embedded in the BR universe, a spacetime immersed in a uniform  Maxwell field. Termed Kerr-Bertotti-Robinson (KBR) black holes, these configurations differ notably from Kerr–Melvin black holes, which are of Petrov type I \cite{Barrientos2024}; KBR spacetimes belong to the Petrov type D class. The Maxwell and Weyl principal null directions in the KBR solution are non-aligned, and
the external field $B$ explicitly deforms the horizon locations and extremality condition \cite{Podolsky2025,Siahaan}. The KBR solution has since been the subject of extensive investigation, encompassing studies on particle dynamics and field perturbations \cite{Gray2026,Mirkhaydarov2601}, magnetic reconnection and the Penrose process \cite{Zeng2025,Mirkhaydarov2601}, as well as light deflection and black hole shadow formation \cite{Mirkhaydarov2601,Ali2026}. 

More recently, the non-twisting subclass of this broader family of type D solutions featuring a non-aligned electromagnetic field was analyzed in \cite{Podolsky2025}. These exact solutions were shown to fall into two principal subclasses. The first describes either the uncharged Schwarzschild black hole or the C-metric embedded in an external BR spacetime, whose geometry is locally $AdS_2 \times S^2$. The second corresponds to a charged Reissner-Nordstr\"{o}m black hole undergoing acceleration in the presence of an external BR electromagnetic field \cite{Podolsky2025}. In the present work, we focus specifically on the uncharged case, namely, the Schwarzschild black hole or the C-metric embedded in a BR background.

It is worth noting that a novel vacuum solution to the Einstein equations has been derived from accelerating electrovacuum black holes embedded in external electromagnetic fields. By applying Melvin–Bonnor magnetization and an inversion symmetry, the external fields are removed, leaving non-trivial gravitational backreactions that result in new accelerating vacuum spacetimes of Petrov type I \cite{Barrientos2026}. The static, non-accelerating limit of the magnetized Schwarzschild solution aligns with known results, while the inversion symmetry yields a genuinely new vacuum configuration. This method provides a systematic way to generate algebraically general vacuum geometries from electromagnetic seeds, highlighting how electromagnetic embeddings can produce non-trivial vacuum metrics.

The Schwarzschild-BR configuration interplays between the black hole mass, the background of a non-aligned magnetic field, and the resulting metric structure, yielding a rich landscape of geodesic phenomena. Building on this work, a further generalization that introduces uniform acceleration into the BR framework was proposed in \cite{Ovcharenk2026}, yielding what may be called the accelerating BR spacetime. This new vacuum solution incorporates the acceleration parameter $\alpha$ familiar from the C-metric an, as special limits, encompasses both the BR spacetime in uniformly accelerating coordinates (the zero-mass limit) and the standard C-metric (when the magnetic field is switched off).

Accelerating spacetimes occupy a distinguished place in general relativity. The C-metric, first written down by Levi-Civita \cite{Levi1918} and later identified as describing a pair of black holes accelerating in opposite directions due to cosmic strings or struts \cite{Kinnersley1970, Bonnor1983}, has been extensively studied over the past decades. Its optical properties, in particular the photon sphere, shadow, and gravitational lensing, have attracted significant attention in the context of present and future very-long-baseline interferometry (VLBI) observations, which now provide resolved images of the shadow of the supermassive black holes M87* \cite{EHTL1} and Sgr A* \cite{EHTL12}. At the same time, the role of external magnetic fields in shaping the orbital dynamics around black holes is well established: magnetic fields affect the location of the ISCO, modify photon trajectories, and leave imprints on the gravitational-wave signal emitted by inspiraling compact objects. A few recent studies of optical properties and particle dynamics around black hole solutions in the presence of a uniform magnetic field were reported in \cite{ Zhang2025, AhmedAlBadawiSakalli2025, Sakalli2025}. 

The accelerating BR spacetime combines two physically significant deformations of the Schwarzschild geometry: a uniform external magnetic field and uniform acceleration. This interplay makes it a useful arena to track, within a single exact vacuum metric, how magnetic confinement and acceleration-induced defocusing compete in shaping orbital stability, strong-field optical observables, and thermodynamic behavior. In this paper we (i) analyze timelike geodesics and equatorial circular motion, deriving the effective potential and closed expressions for $\mathcal{L}_{\rm sp}$ and $\mathcal{E}_{\rm sp}$, and determining the ISCO as a function of $(B,\alpha)$; (ii) go beyond equatorial motion by computing the radial and latitudinal epicyclic frequencies for small harmonic oscillations about circular orbits, which provide a direct diagnostic of local stability against radial and vertical perturbations; (iii) investigate null geodesics and obtain the photon-sphere radius, critical impact parameter, and shadow size, complemented by photon trajectories, the effective radial force, and the Lyapunov exponent controlling the instability of circular null orbits; (iv) present two-dimensional parameter-space surveys for the photon-sphere and shadow observables; and (v) compute the horizon radius and Hawking temperature, clarifying how $\alpha$ affects the surface gravity even though it does not shift the horizon location. Throughout, we verify the recovery of the Schwarzschild, C-metric, and Schwarzschild--BR limits and emphasize the physically distinct roles played by $B$ and $\alpha$.

Throughout this paper we adopt geometrized units $G = c = 1$ and use the metric signature $(-,+,+,+)$.

\section{Spacetime metric}\label{sec2}

The accelerating BR configuration was recently introduced in Ref.~\cite{Podolsky2025}. Before presenting the solution, we recall that exact vacuum metrics sourced by a purely magnetic field with acceleration can be constructed by adapting the Harrison transformation \cite{Harrison1968} and the Ernst solution-generating technique \cite{Ernst1968} to the C-metric seed. The purely magnetic restriction is essential for the consistency of the construction: it guarantees that the twist of the electromagnetic field vanishes in the equatorial plane and simplifies the form of the stress-energy tensor, allowing a clean factorization of the metric functions. 

The resulting line element is \cite{Podolsky2025,Ovcharenk2026}
\begin{align}
\mathrm{d} s^2=\frac{1}{\Omega^2}\Big[-\frac{Q}{r^2} \mathrm{d} t^2&+\frac{r^2\mathrm{d} r^2}{Q}\notag\\&+r^2\left(\frac{\mathrm{d} \theta^2}{P}+P \sin ^2 \theta \mathrm{d} \varphi^2\right)\Big] ,\label{aa1}
\end{align}
where
\begin{align}
P & =1-2 \alpha m \cos \theta+B^2 m^2 \cos ^2 \theta,\nonumber\\
Q & =\left(r^2-B^2 m^2 r^2-2 m r \right)\left(1+\left(B^2-\alpha^2\right) r^2\right),\nonumber\\
\Omega^2 & =(1-\alpha r \cos \theta)^2\notag\\&+B^2\left[r^2\left(P-\cos ^2 \theta\right)+2 m r \cos ^2 \theta\right], \label{aa2}
\end{align}
and the gauge field
\begin{equation}
A_{\mu} dx^{\mu}=\frac{1}{2}\left(r\,\Omega_{,r}-\Omega+1\right)\,d\varphi.\label{aa3}
\end{equation}

Here, we restrict attention to the purely magnetic case, since in this setting the construction of the new vacuum solution becomes possible. The relevant limiting procedures have already been analyzed in \cite{Ovcharenk2026}, thereby justifying the description of the geometry as an accelerating BR spacetime. In the zero-mass limit, the solution reduces to the BR background in uniformly accelerating coordinates, whereas switching off the magnetic field recovers the standard C-metric. Conversely, vanishing acceleration leads to the Schwarzschild-BR configuration discussed in \cite{Podolsky2025}, which is characterized by
\begin{align}
P & =1+B^2 m^2 \cos ^2 \theta,\nonumber\\
Q & =\left(r-B^2 m^2 r^2-2 m r\right)\left(1+B^2 r^2\right),\nonumber\\
\Omega^2 & =1+B^2\left[r^2 \sin ^2 \theta+\left(2 m r+B^2 m^2 r^2\right) \cos ^2 \theta\right],\label{aa4}
\end{align}
with the same gauge field \eqref{aa3}.

As usual, the acceleration is provided by a conical defect; in this case, 
\begin{equation}
  \lim_{\theta\rightarrow 0,\pi}\frac{1}{\sin\theta}\int_0^{2\pi}\sqrt{\frac{g_{\varphi\varphi}}{g_{\theta\theta}}}\mathrm{d}\varphi=2\pi(1\mp2\alpha m+B^2m^2).\label{aa5}
\end{equation}

This shows that the conical singularity cannot be removed at both poles simultaneously unless $\alpha m$ and $B^2 m^2$ satisfy a special relation. In practice, one typically removes the deficit on one side (say $\theta=\pi$, the south pole) by an appropriate choice of the period of $\varphi$, leaving the string or strut that provides the acceleration attached at the other pole.

Analysis of the radial null geodesics in this space-time yields (setting $ds^2=0$, $d\theta=0=d\phi$)
\begin{equation}
    \frac{dr}{dt}=\frac{Q}{r^2}.\label{aa6}
\end{equation}
This define a coordinate speed light for the metric (\ref{aa1}) as,
\begin{equation}
    v(r)=\left|\frac{dr}{dt}\right|=r^2\left|1-m^2 B^2-\frac{2 m}{r}\right|\left|\frac{1}{r^2}+B^2-\alpha^2\right|.\label{aa7}
\end{equation}
Hence, one finds the refractive index given by
\begin{equation}
    n(r)=\frac{c}{r^2}\,\left|1-m^2 B^2-\frac{2 m}{r}\right|^{-1}\left|\frac{1}{r^2}+B^2-\alpha^2\right|^{-1},\label{aa8}
\end{equation}
where $c$ being the speed of light in vacuum and considered it to be unity.

\begin{figure}[ht!]
\centering
\includegraphics[width=0.95\linewidth]{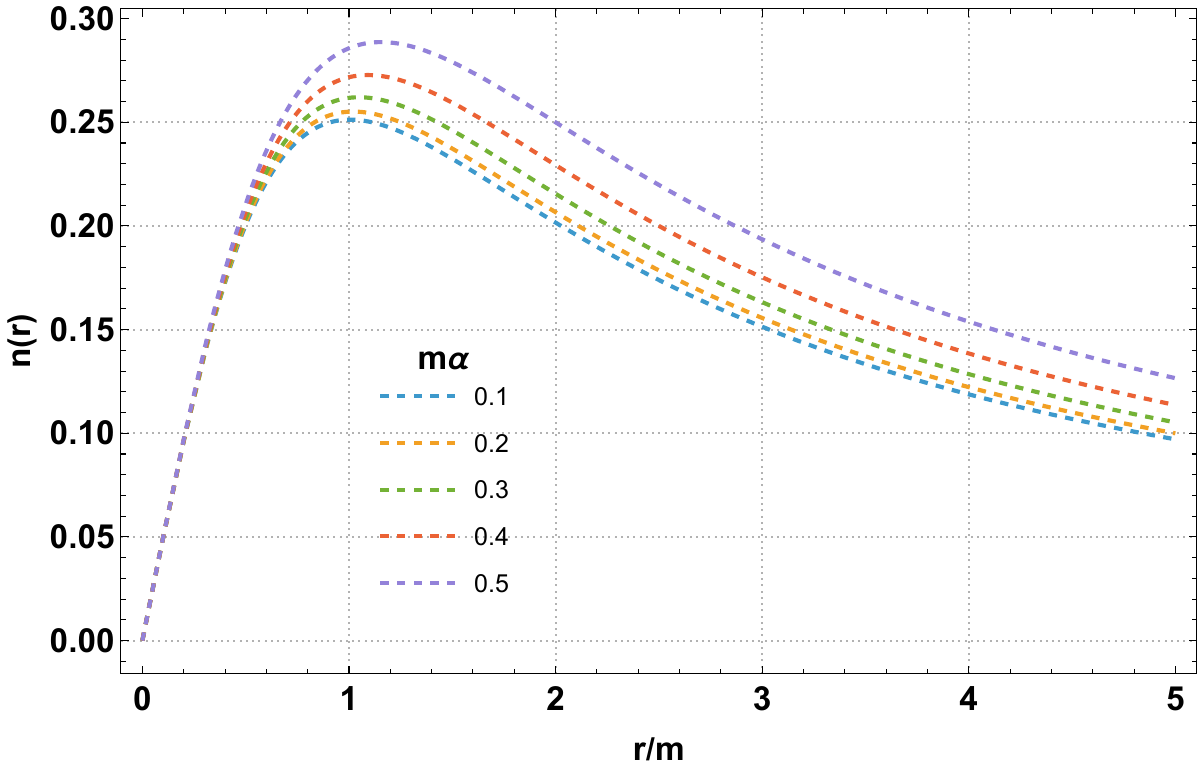}\\
(a) $B=1/m$\\
\includegraphics[width=0.95\linewidth]{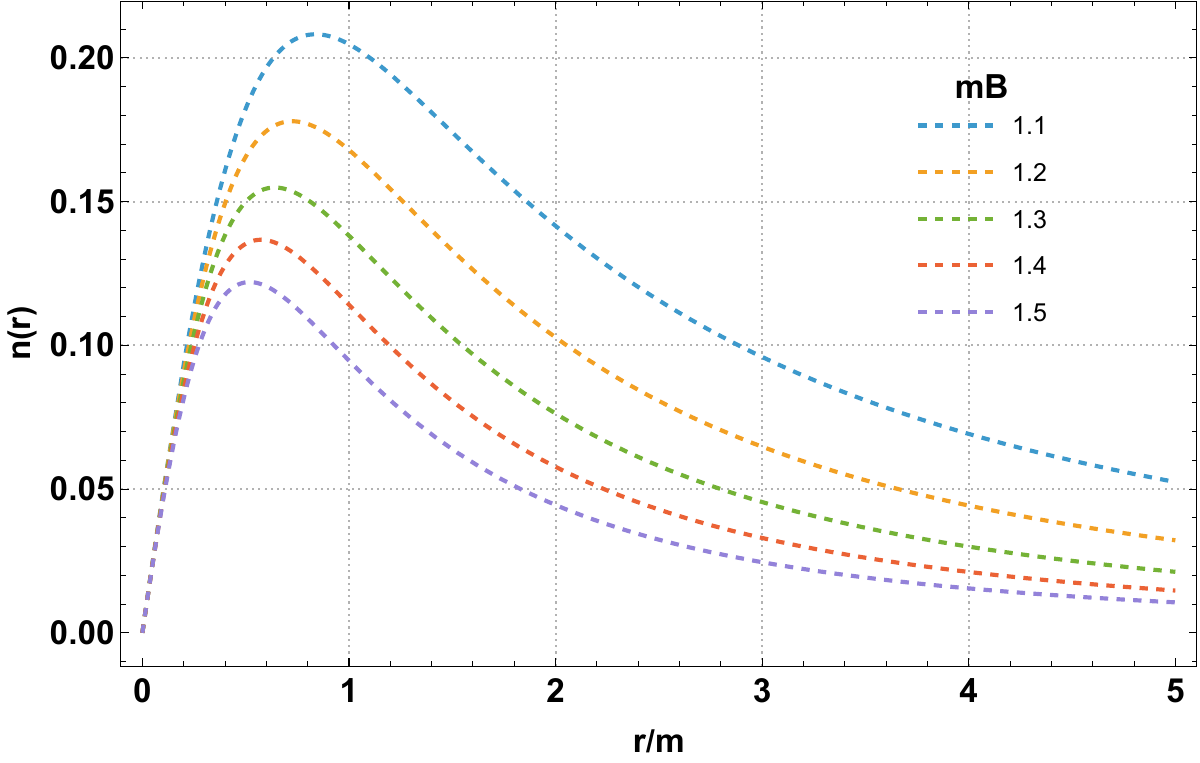}\\
(b)$\alpha=0.01/m$
\caption{The behavior of the refractive index as a function $r/m$ by varying $B$ and $\alpha$.}
\label{fig:refractive}
\end{figure}

In Fig.~\ref{fig:refractive}, we illustrate the variation of the refractive index as a function of the dimensionless radial coordinate $r/m$ for different values of the magnetic field $B$ and the acceleration parameter $\alpha$.

Panel~\ref{fig:refractive}(a) shows that the refractive index initially increases with $r/m$, reaches a maximum value at a certain radial distance, and then gradually decreases as $r/m$ continues to increase. Moreover, the height of this peak increases with increasing values of the acceleration parameter $\alpha$. This indicates that stronger acceleration enhances the maximum refractive behavior of the medium without altering the overall trend of variation.

In contrast, panel~\ref{fig:refractive}(b) exhibits a similar qualitative behavior: the refractive index rises with $r/m$, attains a peak, and then decreases. However, in this case, increasing the magnetic field $B$ reduces the magnitude of the peak. Thus, while both parameters preserve the general profile of the refractive index, they influence its maximum value in opposite ways-$\alpha$ amplifies the peak, whereas $B$ suppresses it.

\begin{figure}[ht!]
\centering
\includegraphics[width=\linewidth]{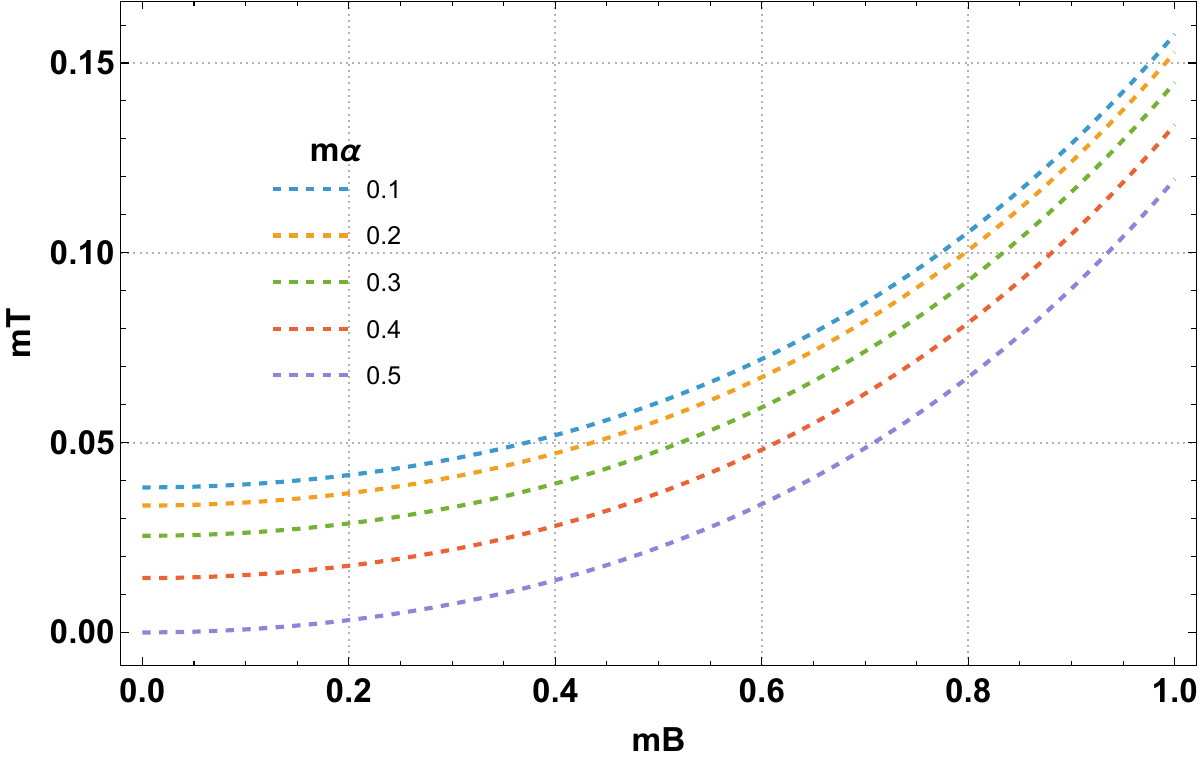}\\
(a)\\
\includegraphics[width=\linewidth]{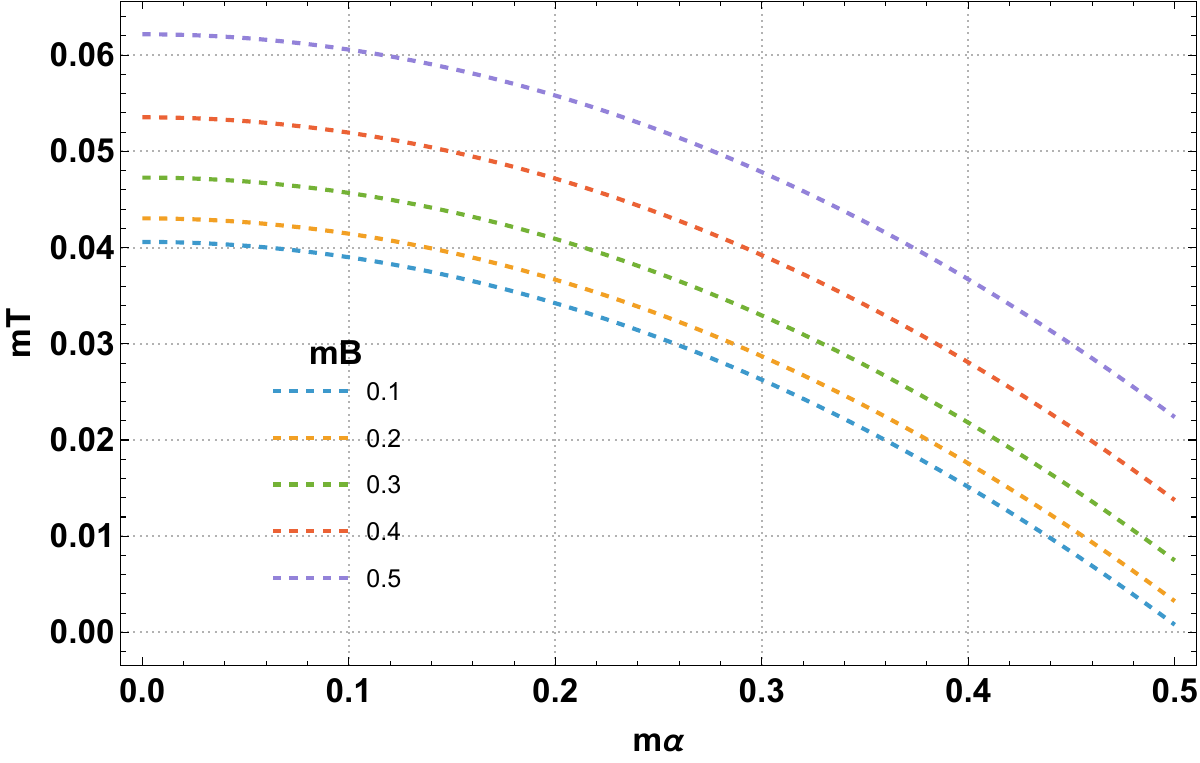}\\
(b)
\caption{The behavior of the Hawking temperature as a function of the magnetic field $B$ and acceleration $\alpha$.}
\label{fig:temeprature}
\end{figure}

\section{Hawking Temperature}

The Hawking temperature is the temperature associated with black holes due to quantum effects near the event horizon, predicted by Stephen Hawking  \cite{Hawking1975}. It implies that black holes emit thermal radiation (Hawking radiation) with a temperature $T$ related to the surface gravity $\kappa$ by $T=\kappa/2\pi$ \cite{Hawking1975,Wald1994}.

The horizon radius can be determined using the condition $Q(r_h)=0$, where $Q(r)$ is given in (\ref{aa2}). We find
\begin{equation}
    [r_h\,(1-m^2 B^2)-2 m] r_h=0\Rightarrow r_h=\frac{2 m}{1-m^2 B^2}.\label{zz1}
\end{equation}
Noted that the horizon exist provided we have $m^2 B^2 <1$.

The surface gravity at the horizon, in our case, is defined by
\begin{equation}
    \kappa=-\frac{1}{2} \frac{d}{dr} g^{\rm eff}_{tt}\Big{|}_{r=r_h}.\label{zz2}
\end{equation}
For our metric, $g^{\rm eff}_{tt}=-Q(r)/r^2$, so the surface gravity is
\begin{equation}
    \kappa=\frac{Q'(r_h)}{2 r_h^2}.\label{zz3}
\end{equation}
Substituting $Q(r)$, one finds
\begin{equation}
    \kappa=m \left(\frac{1}{r_h^2}+B^2-\alpha^2\right).\label{zz4}
\end{equation}

Finally, the Hawking temperature reads
\begin{align}
    T&=\frac{2 m [1+(B^2-\alpha^2)\,r_h^2]}{4 \pi r_h^2}\nonumber\\
    &=\frac{(1-m^2 B^2)^2+4 m^2 (B^2-\alpha^2)}{8\pi m}.\label{zz5}
\end{align}

Below, we present some special cases.
\begin{itemize}
    \item When $\alpha=0$, corresponding to the absence of acceleration, the Hawking temperature simplifies as
    \begin{equation}
        T=\frac{(1+m^2 B^2)^2}{8\pi m}.\label{zz6}
    \end{equation}

    \item When $B=0$, corresponding to the absence of a magnetic field, the Hawking temperature simplifies as
    \begin{equation}
        T=\frac{1-4 m^2 \alpha^2}{8\pi m}.\label{zz7}
    \end{equation}
    \item When $\alpha=0$ and $B=0$, the Hawking temperature simplifies as
    \begin{equation}
        T=\frac{1}{8\pi m},\label{zz8}
    \end{equation}
    which is similar to the Schwarzschild black hole case.
\end{itemize}

From the above analysis, we observe that although the location of the event horizon is independent of the parameter $\alpha$, the Hawking temperature (\ref{zz5}) explicitly depends on both $\alpha$ and the magnetic field $B$. This indicates that while $\alpha$ does not modify the horizon radius, it influences the surface gravity, thereby affecting the black hole’s thermal properties. 

In Fig.~\ref{fig:temeprature}, we present the behavior of the Hawking temperature as a function of the magnetic field $B$ and the acceleration parameter $\alpha$. Panel~\ref{fig:temeprature}(a) shows that the temperature increases with increasing $B$, exhibiting a quadratic (parabolic) dependence. This indicates that the magnetic field enhances the black hole's thermal radiation. In contrast, panel~\ref{fig:temeprature}(b) demonstrates that the Hawking temperature decreases as $\alpha$ increases, implying that acceleration suppresses the thermal emission.

\section{Dynamics of Test Particles}\label{sec3}

In this section, we study the dynamics of massive test particles in the equatorial plane of the accelerating BR spacetime and analyze how the magnetic field strength $B$ and the acceleration parameter $\alpha$ affect the orbital structure. The main quantities of interest are the effective potential, the specific orbital energy and angular momentum for circular motion, and the radius of the innermost stable circular orbit (ISCO). These quantities encode important information about the accretion structure around the black hole and serve as direct observables in high-energy astrophysical scenarios involving magnetized environments. For the Schwarzschild black hole, the ISCO radius will be $r_{\rm ISCO}=6 M$ \cite{chandrasekhar1984}.

\begin{figure*}[ht!]
\centering
\includegraphics[width=\textwidth]{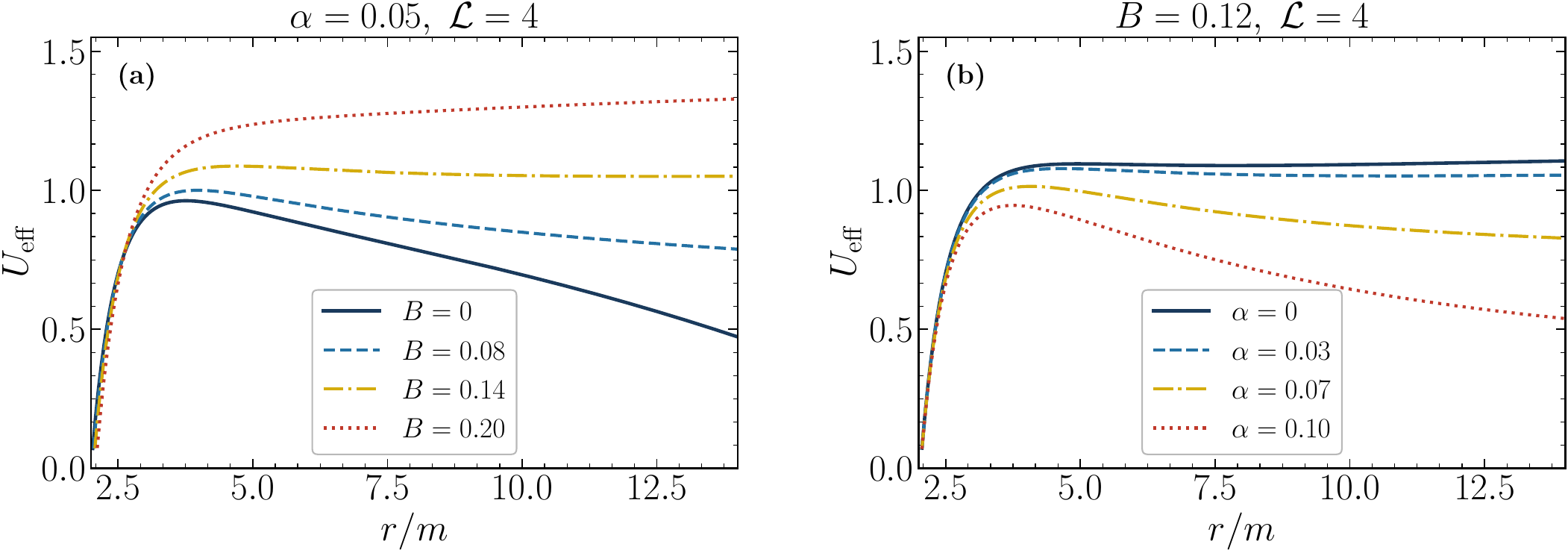}
\caption{Effective potential $U_{\rm eff}(r)$ for massive test particles [Eq.~(\ref{bb6})] as a function of the radial coordinate $r/m$, with $m=1$ and angular momentum $\mathcal{L}=4$. Panel~(a) shows the effect of varying the magnetic field strength $B \in \{0, 0.08, 0.14, 0.20\}$ at fixed acceleration $\alpha=0.05$. Panel~(b) shows the effect of varying the acceleration parameter $\alpha \in \{0, 0.03, 0.07, 0.10\}$ at fixed $B=0.12$. In panel~(a), increasing $B$ systematically raises the height of the potential barrier and shifts its local maximum inward, reflecting the magnetic confinement that tightens circular orbits. In panel~(b), increasing $\alpha$ lowers and broadens the barrier, indicating that acceleration reduces the orbital binding energy. The standard Schwarzschild effective potential is recovered in the $B=\alpha=0$ limit (solid dark curve in panel~(a)). The vertical dashed lines at small $r$ indicate the respective event-horizon radii $r_h = 2m/(1-B^2m^2)$ for each value of $B$.}
\label{fig:1}
\end{figure*}

The Lagrangian density function for a test particle of mass $\mu$ in curved space-time is given by \cite{chandrasekhar1984,Wald1984} 
\begin{equation}
    \mathrm{L}=\frac{1}{2}\,\mu\,g_{\mu\nu}\,\frac{dx^{\mu}}{d\lambda}\,\frac{dx^{\nu}}{d\lambda},\label{bb1}
\end{equation}
where $\lambda$ is the affine parameter (proper time for massive particles, which we normalize so that $g_{\mu\nu}\dot{x}^\mu\dot{x}^\nu = -1$).

Restricting to the equatorial plane, $\theta=\pi/2$, which is a symmetry plane of the metric since all metric functions are even in $(\theta - \pi/2)$ on that surface, the Lagrangian (\ref{bb1}) simplifies using the metric (\ref{aa1}) to
\begin{widetext}
\begin{align}
    \mathrm{L}=\frac{\mu}{2}\,\Bigg[-\frac{Q}{r^2 (1+B^2 r^2)} \left(\frac{dt}{d\lambda}\right)^2+\frac{r^2}{Q\,(1+B^2 r^2)}\,\left(\frac{dr}{d\lambda}\right)^2+\frac{r^2}{(1+B^2 r^2)} \left(\frac{d\varphi}{d\lambda}\right)^2\Bigg],\label{bb2}
\end{align}
\end{widetext}
where we have used $P\big|_{\theta=\pi/2} = 1$ and $\Omega^2\big|_{\theta=\pi/2} = 1 + B^2 r^2 \sin^2(\pi/2) = 1 + B^2 r^2$.

There exist two conserved quantities associated with the Killing vectors $\xi^{\mu}_{(t)} \equiv \partial_t$ (temporal stationarity) and $\xi^{\mu}_{(\varphi)} \equiv \partial_{\varphi}$ (axial symmetry). These are the conserved energy and angular momentum per unit mass:
\begin{align}
    \mathcal{E}&=\frac{Q}{r^2 (1+B^2 r^2)} \left(\frac{dt}{d\lambda}\right),\label{bb3}\\
    \mathcal{L}&=\frac{r^2}{(1+B^2 r^2)} \left(\frac{d\varphi}{d\lambda}\right).\label{bb4}
\end{align}

\begin{figure*}[tbhp]
\centering
\includegraphics[width=\textwidth]{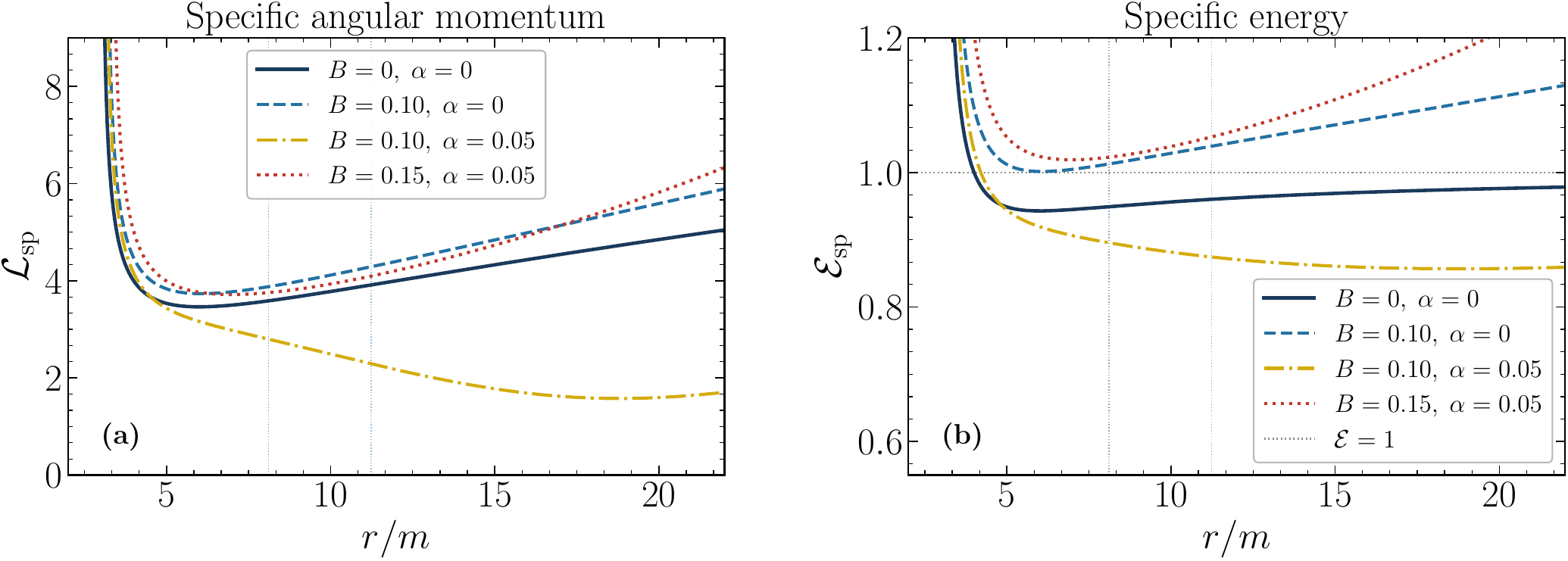}
\caption{Specific angular momentum $\mathcal{L}_{\rm sp}$ (panel~a, Eq.~(\ref{bb8})) and specific energy $\mathcal{E}_{\rm sp}$ (panel~b, Eq.~(\ref{bb10})) for circular orbits as functions of the radial coordinate $r/m$, shown for four representative parameter combinations: $(B,\alpha) = (0,0)$ (Schwarzschild, solid dark), $(0.10, 0)$, $(0.10, 0.05)$, and $(0.15, 0.05)$. The dotted vertical lines mark the corresponding ISCO radii obtained numerically from the marginal stability condition [Eq.~(\ref{bb11})]; the minimum of each $\mathcal{L}_{\rm sp}$ curve coincides with the respective ISCO. The horizontal dotted line in panel~(b) marks the marginal energy $\mathcal{E}=1$ separating bound ($\mathcal{E}<1$) from unbound ($\mathcal{E}>1$) circular orbits. Increasing $B$ generally raises the angular momentum curve and shifts the ISCO to larger radii, while increasing $\alpha$ at fixed $B$ lowers both $\mathcal{L}_{\rm sp}$ and $\mathcal{E}_{\rm sp}$. The divergence of $\mathcal{L}_{\rm sp}$ near the photon sphere and its minimum at the ISCO are characteristic features of every Schwarzschild-like geometry; the parameters $B$ and $\alpha$ control the location and height of these features.}
    \label{fig:2}
\end{figure*}

Using the normalization condition $g_{\mu\nu}\,\dot{x}^{\mu}\dot{x}^{\nu}=-1$ together with Eqs.~(\ref{bb3})--(\ref{bb4}), we arrive at the radial equation of motion
\begin{equation}
    \frac{1}{(1+B^2 r^2)^2}\,\left(\frac{dr}{d\lambda}\right)^2+U_{\rm eff}(r)=\mathcal{E}^2,\label{bb5}
\end{equation}
where the effective potential of the system is
\begin{widetext}
\begin{align}
    U_{\rm eff}(r)=\frac{1+(B^2-\alpha^2) r^2}{1+B^2 r^2}\left[1+\left(B^2+\frac{1}{r^2}\right)\,\mathcal{L}^2\right]\,f(r),\qquad
    f(r)=1-\frac{2 m}{r}-m^2 B^2\label{bb6}
\end{align}
\end{widetext}

The behavior of $U_{\rm eff}$ is illustrated in Fig.~\ref{fig:1}. In Fig. \ref{fig:1}(a), one clearly sees that increasing the magnetic field $B$ raises the potential barrier and moves its peak closer to the horizon, while in Fig. \ref{fig:1}(b) the acceleration $\alpha$ acts in the opposite sense, reducing the barrier height. In the limit $B \to 0$, $\alpha \to 0$, the standard Schwarzschild effective potential is recovered. The competition between these two effects is physically significant: for sufficiently large $B$, the potential barrier can lie well above the Schwarzschild value, whereas large $\alpha$ can suppress it entirely, destabilizing circular orbits at comparatively small radii.

For circular orbits, the following conditions must be satisfied:
\begin{equation}
U_{\rm eff}(r)=\mathcal{E}^2,\quad \frac{d U_{\rm eff}}{dr}=0.\label{bb7} 
\end{equation}

Solving the second condition with the effective potential in Eq.~(\ref{bb6}), we find the specific angular momentum for circular orbits:
\begin{equation}
\mathcal{L}^2_{\rm sp}=\frac{H(r)\, r^3}{\,2-r\,H(r)\, \left(1+r^2\,B^2\right)},\label{bb8}
\end{equation}
where 
\begin{align}
H(r)&=\frac{2r(B^2-\alpha^2)}{1+(B^2-\alpha^2)r^2}-\frac{2rB^2}{1+B^2 r^2}\notag\\&+\frac{2m/r^2}{1-\frac{2 m}{r}-m^2 B^2}.\label{bb9}
\end{align}
Using Eqs.~(\ref{bb7}) and (\ref{bb8}), the specific energy for circular orbits is:
\begin{widetext} 
\begin{equation}
\mathcal{E}^2_{\rm sp}=\frac{2\,\left(1-\frac{2 m}{r}-m^2 B^2\right)\,\big[1+(B^2-\alpha^2) r^2\big]
}{\big(1+B^2 r^2\big)\left[2-r\!\left(1+r^2\,B^2\right)
\!\left(\frac{2r(B^2-\alpha^2)}{1+(B^2-\alpha^2)r^2}-
\frac{2rB^2}{1+B^2 r^2}+\frac{2m/r^2}{1-\frac{2 m}{r}-m^2 B^2}\right)\right]}.\label{bb10}
\end{equation}
\end{widetext}
From the above expressions, we observe that the specific angular momentum and specific energy of test particles revolving in circular orbits are influenced by the magnetic field strength $B$, the acceleration parameter $\alpha$, and the black hole mass $m$. In the Schwarzschild limit ($B=\alpha=0$), one recovers the well-known results $\mathcal{L}^2_{\rm sp} = mr^2/(r-3m)$ and $\mathcal{E}^2_{\rm sp} = (r-2m)^2/[r(r-3m)]$.

The radial profiles of $\mathcal{L}_{\rm sp}$ and $\mathcal{E}_{\rm sp}$ are shown in Fig.~\ref{fig:2} for four representative combinations of $(B,\alpha)$. Each curve exhibits a well-defined minimum that identifies the ISCO, where circular motion transitions from stable (minimum) to unstable (inward). Figure \ref{fig:2}(a) shows that larger $B$ requires more angular momentum to sustain a circular orbit at a given radius, consistent with the heightened potential barrier seen in Fig.~\ref{fig:1}(a). Figure \ref{fig:2}(b) reveals that, for the same parameter sets, the specific energy curves all lie close to unity near the ISCO, as expected from the near-marginal binding of these orbits. The dotted vertical lines indicate the numerically computed ISCO radii; their outward displacement with increasing $B$ and inward displacement with increasing $\alpha$ are evident.

For stable circular orbits, the additional condition
\begin{equation}
\frac{d^2U_{\rm eff}}{dr^2}\geq 0\label{bb11}
\end{equation}
must be satisfied. The ISCO radius $r_{\rm ISCO}$ is determined by the marginal stability condition $d^2U_{\rm eff}/dr^2 = 0$, which, combined with conditions (\ref{bb7}), yields the simultaneous system
\begin{align}
&\frac{C'}{C}+\frac{D'}{D}+\frac{f'(r)}{f(r)}=0,\nonumber\\
\frac{C''}{C}-\left(\frac{C'}{C}\right)^2+\frac{D''}{D}&-\left(\frac{D'}{D}\right)^2
\notag\\&+\frac{f''(r)}{f(r)}-\left(\frac{f'(r)}{f(r)}\right)^2=0,\label{bb12}
\end{align}
where
\begin{align}
    &C(r)=\frac{1+(B^2-\alpha^2)r^2}{1+B^2 r^2},\nonumber\\
    &D(r)=1+\left(B^2+\frac{1}{r^2}\right)\mathcal{L}^2_{\rm sp}(r).\label{bb13}
\end{align}
This system must, in general, be solved numerically.

\begin{figure}[t]
    \centering
    \includegraphics[width=\columnwidth]{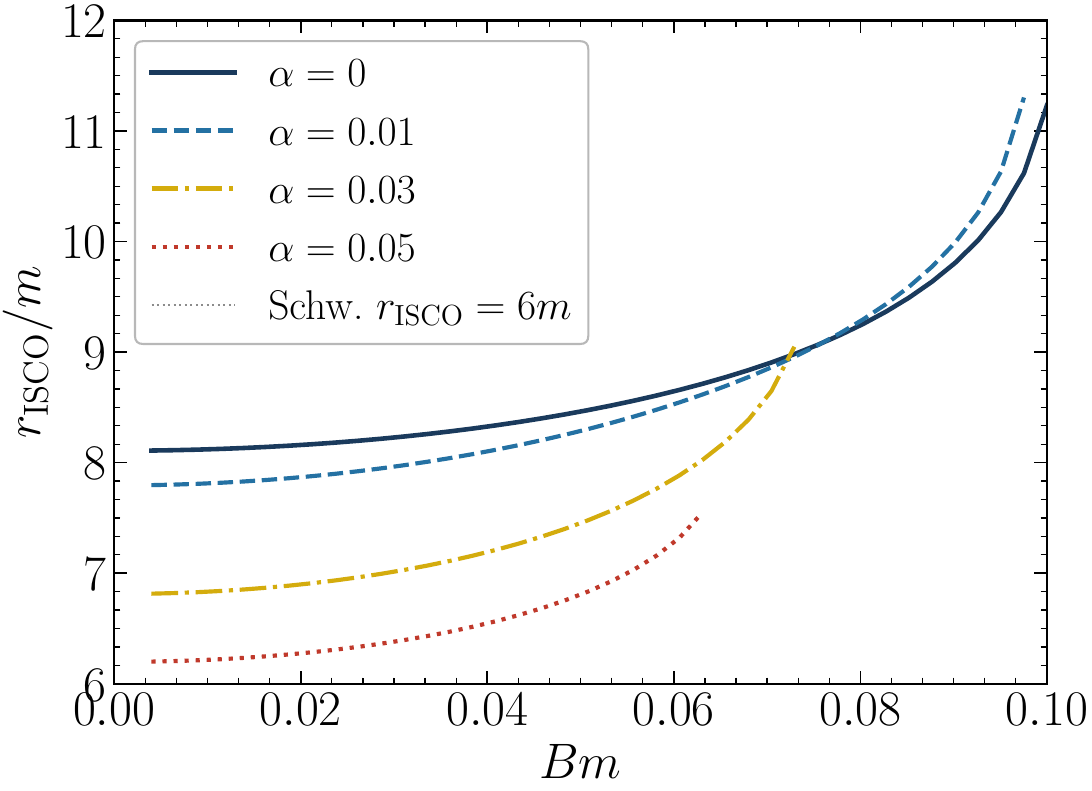}
    \caption{ISCO radius $r_{\rm ISCO}/m$ as a function of the magnetic field parameter $Bm$ for four values of the acceleration parameter: $\alpha \in \{0, 0.01, 0.03, 0.05\}$. The dotted horizontal line marks the Schwarzschild reference value $r_{\rm ISCO} = 6m$. Each curve was obtained by numerically solving the marginal stability condition $d^2U_{\rm eff}/dr^2=0$ along the family of circular orbits parameterized by $L_{\rm sp}(r)$. For fixed $\alpha$, increasing $B$ moves the ISCO to larger radii; the curves terminate at the maximum $B$ for which a stable circular orbit can exist outside the horizon, beyond which no ISCO is found in the scanned range. For fixed $B$, increasing $\alpha$ shifts the entire curve downward, indicating that acceleration counteracts the magnetic confinement and can restore stability at smaller radii. At small $B$ and $\alpha=0$, all curves start above $6m$ because the conformal factor $\Omega^2$ modifies the effective potential even at leading order in $B$; only in the strict double limit $B,\alpha\to 0$ is the Schwarzschild value recovered.}
    \label{fig:3}
\end{figure}

The ISCO is sensitive to both $B$ and $\alpha$: the magnetic field tends to shift $r_{\rm ISCO}$ outward (away from the black hole), while acceleration counteracts this effect for the parameter ranges considered here. These competing tendencies are illustrated clearly in Fig.~\ref{fig:3}, which shows $r_{\rm ISCO}$ as a function of $Bm$ for several values of $\alpha$. The qualitative behavior is similar to that found in related magnetized C-metric spacetimes, but the quantitative values are modified by the BR background.

Finally, we analyze the equatorial trajectories of massive test particles and explicitly show how the parameters $(B,\alpha)$ deform the orbital paths relative to the Schwarzschild baseline. Combining the conserved angular momentum \eqref{bb4} with the radial energy-balance equation \eqref{bb5}, and eliminating the affine parameter, we obtain the orbit equation in the form
\begin{align}
    \left(\frac{1}{r^2}\frac{dr}{d\varphi}\right)^2
   & =\frac{\mathcal{E}^2}{\mathcal{L}^2}
   \notag\\& -\frac{1+(B^2-\alpha^2) r^2}{1+B^2 r^2}
    \left[\frac{1}{\mathcal{L}^2}+B^2+\frac{1}{r^2}\right]\,f(r),
    \label{kk1}
\end{align}
where $f(r)=1-\dfrac{2m}{r}-m^2B^2$. Introducing the inverse radius $u(\varphi)=1/r(\varphi)$, Eq.~\eqref{kk1} becomes
\begin{equation}
    \left(\frac{du}{d\varphi}\right)^2
    =\frac{\mathcal{E}^2}{\mathcal{L}^2}
    -\frac{u^2+B^2-\alpha^2}{u^2+B^2}
    \left[\frac{1}{\mathcal{L}^2}+B^2+u^2\right]\,f(u),
    \label{kk2}
\end{equation}
with $f(u)=1-2mu-m^2B^2$. Equation~\eqref{kk2} is particularly useful because it makes the orbit classification transparent: the right-hand side must be non-negative along the motion, and its zeros determine the radial turning points. In particular, one turning point corresponds to a scattering orbit (a single closest-approach radius), while two turning points correspond to a bound orbit (radial oscillation between periapsis and apoapsis).

\begin{itemize}
\item \textit{Vanishing acceleration} ($\alpha=0$). In this case, Eq.~\eqref{kk2} simplifies to
\begin{align}
\left(\frac{du}{d\varphi}\right)^2+(1-m^2 B^2)u^2
&=\frac{\mathcal{E}^2}{\mathcal{L}^2}
-\left(\frac{1}{\mathcal{L}^2}+B^2\right)(1-m^2 B^2)
\nonumber\\
&\quad+2m\left(\frac{1}{\mathcal{L}^2}+B^2\right)u+2mu^3,
\label{kk3}
\end{align}
and differentiating with respect to $\varphi$ yields the corresponding second-order form
\begin{align}
\frac{d^2u}{d\varphi^2}+(1-m^2 B^2)u
= m\left(\frac{1}{\mathcal{L}^2}+B^2\right)+3mu^2.
\label{kk4}
\end{align}

\item \textit{Vanishing magnetic field} ($B=0$). Equation~\eqref{kk2} reduces to the acceleration-only orbit equation
\begin{equation}
\left(\frac{du}{d\varphi} \right)^2 
= \frac{\mathcal{E}^2}{\mathcal{L}^2} 
- \left(1 - \frac{\alpha^2}{u^2}\right)\left(\frac{1}{\mathcal{L}^2} +u^2\right) 
\left( 1 - 2 m u \right),
\label{kkB0}
\end{equation}
which shows explicitly how $\alpha$ reshapes the effective balance between the centrifugal and attractive terms.

\item \textit{Schwarzschild limit} ($B=\alpha=0$). One recovers the standard Schwarzschild orbit equation
\begin{align}
\left(\frac{du}{d\varphi}\right)^2+u^2
=\frac{\mathcal{E}^2-1}{\mathcal{L}^2}+\frac{2m}{\mathcal{L}^2}u+2mu^3,
\label{kk5}
\end{align}
or, equivalently, its familiar second-order form
\begin{align}
\frac{d^2u}{d\varphi^2}+u=\frac{m}{\mathcal{L}^2}+3mu^2.
\label{kk6}
\end{align}
\end{itemize}

\begin{figure*}[tbhp]
\centering
\includegraphics[scale=0.5]{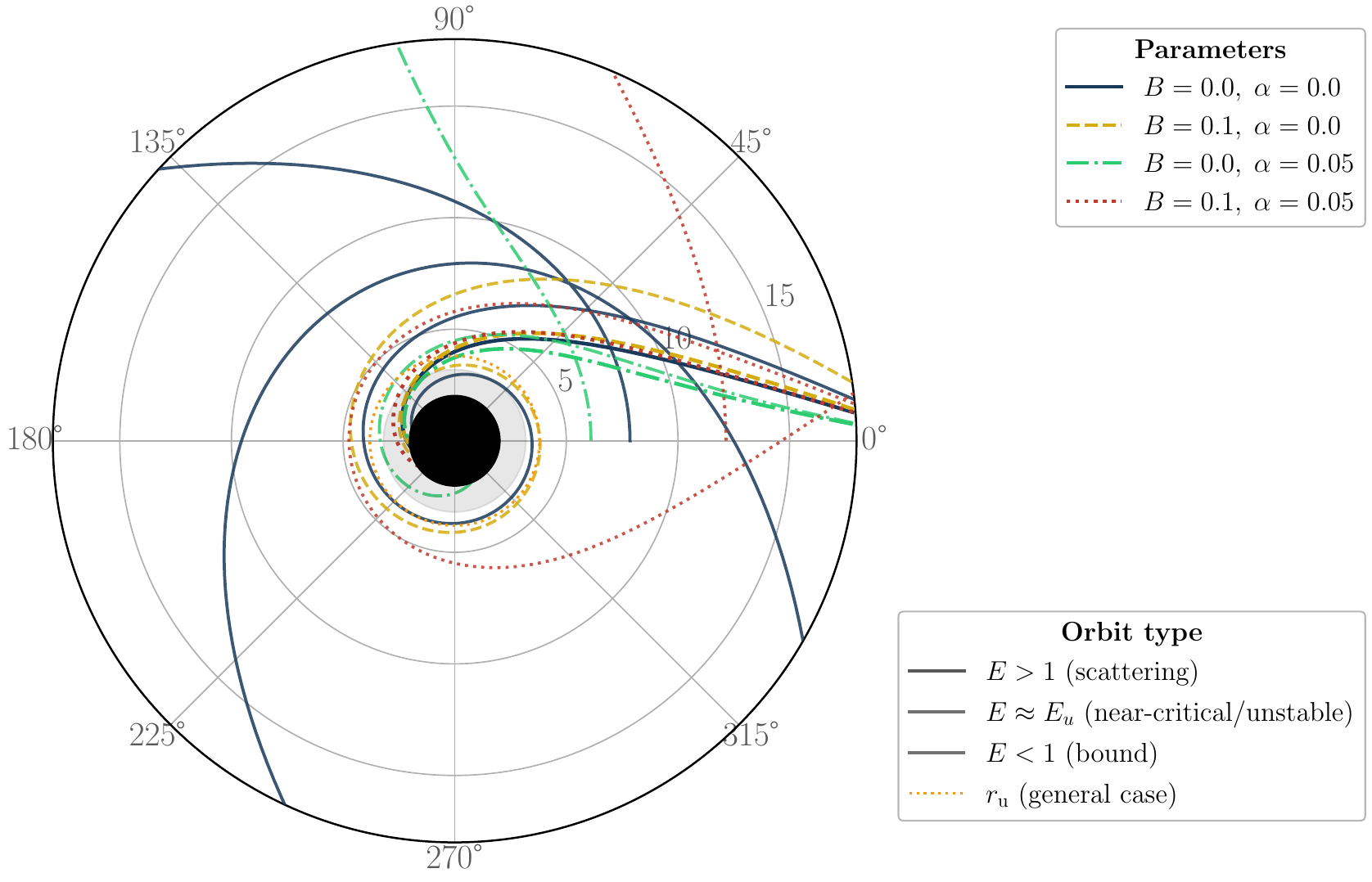}
\caption{Representative massive-particle trajectories in the equatorial plane of the accelerating BR spacetime obtained from the orbit equation \eqref{kk1} and its inverse-radius form \eqref{kk2}. The angular momentum is fixed to $\mathcal{L}=4.5\,m$ and the trajectories start at $r_{\rm start}=22\,m$. For each parameter pair $(B,\alpha)$ in the legend, three energies are chosen to illustrate distinct regimes: a scattering orbit with $\mathcal{E}>\mathcal{E}_u$ (above the potential barrier), a near-critical orbit with $\mathcal{E}\approx\mathcal{E}_u$ (close to the unstable circular orbit at $r=r_u$), and a bound orbit with $\mathcal{E}<1$ (two turning points). Numerically, we integrate the second-order equation obtained by differentiating \eqref{kk2} with respect to $\varphi$ and dividing by $2\,du/d\varphi$, i.e.\ $u''(\varphi)=-\frac{1}{2}\,\frac{d}{du}[C(u)D(u)f(u)]$ with $C(u)=(u^2+B^2-\alpha^2)/(u^2+B^2)$ and $D(u)=1/\mathcal{L}^2+B^2+u^2$; the energy $\mathcal{E}$ enters through the initial condition for $du/d\varphi$ determined from \eqref{kk2}. The black disk represents the region $r<r_h$, and the orange dotted circle marks the unstable circular-orbit radius $r_u$ for the general case $(B,\alpha)=(0.10,0.05)$.}
\label{fig:orbits_eq19}
\end{figure*}

Figure~\ref{fig:orbits_eq19} provides a geometric counterpart of the effective-potential analysis discussed above. For fixed $(B,\alpha)$ and fixed $\mathcal{L}$, the equatorial effective potential $U_{\rm eff}(r)$ in Eq.~\eqref{bb6} typically exhibits a local maximum (unstable circular orbit) and, when stable circular motion exists, a local minimum (stable circular orbit). Denoting by $\mathcal{E}_u^2$ the value of $U_{\rm eff}$ at its local maximum, the orbit families in Fig.~\ref{fig:orbits_eq19} can be interpreted as follows.
(i) For $\mathcal{E}>\mathcal{E}_u$, the particle can overcome the potential barrier: it reaches a minimum radius $r_{\min}$ where the right-hand side of \eqref{kk2} vanishes and then scatters back to large radii, producing a deflected (open) trajectory.
(ii) For $\mathcal{E}\approx\mathcal{E}_u$, the motion is tuned close to the separatrix: the radial evolution becomes very slow near the unstable circular orbit at $r=r_u$, so the trajectory exhibits a pronounced ``whirl'' phase around the black hole before escaping or plunging inward depending on how $\mathcal{E}$ is tuned relative to $\mathcal{E}_u$. (iii) For $\mathcal{E}<1$ and when two turning points exist, the motion is bound: $u(\varphi)$ oscillates between two roots of \eqref{kk2}, generating a rosette-like orbit confined between periapsis and apoapsis. The black disk indicates capture ($r<r_h$), emphasizing that trajectories with sufficiently small turning points ultimately cross the horizon.

The dependence on $(B,\alpha)$ follows directly from the structure of \eqref{kk1}--\eqref{kk2}. The magnetic field $B$ affects both the redshift factor $f$ and the rational prefactors multiplying it, thereby shifting the location of the horizon $r_h=2m/(1-B^2m^2)$ and modifying the height and position of the effective barrier. In contrast, $\alpha$ enters through the combination $(B^2-\alpha^2)$ and through the factor $(u^2+B^2-\alpha^2)/(u^2+B^2)=1-\alpha^2/(u^2+B^2)$, which reshapes the balance between the centrifugal contribution and the effective attraction at a given $u$. Consequently, varying $(B,\alpha)$ shifts the turning points of \eqref{kk2} and the separatrix between scattering, bound, and capture trajectories, yielding orbit deformations that are not captured by the Schwarzschild limit alone.

For reference, Eq.~\eqref{kk2} reproduces the expected special limits: setting $\alpha=0$ yields Eqs.~\eqref{kk3}--\eqref{kk4}; setting $B=0$ yields Eq.~\eqref{kkB0}; and in the double limit $B=\alpha=0$ one recovers the standard Schwarzschild relations \eqref{kk5}--\eqref{kk6}.

\subsection{Advance of the Perihelion}
In this section, we solve the trajectory equation (\ref{kk4}) for zero acceleration parameter $a=0$ showing how the magnetic field alter the the perihelion shifts for the orbit of the planets.

To solve Eq.~(\ref{kk4}) perturbatively, and noting that $m B \ll 1$, it is valid to treat the last term as a relativistic correction compared with the Newtonian case. The perturbative solution is expressed in terms of a small parameter $\epsilon = 3M^2/\mathcal{L'}^2$, where $\frac{1}{\mathcal{L'}^2}=\frac{1}{\mathcal{L}^2}+B^2$ as:
\begin{equation}
u \simeq u^{(0)} + \epsilon u^{(1)}, \label{zz1}
\end{equation}
where $u^{(0)}$ satisfies the zeroth-order differential equation
\begin{equation}
\frac{d^2u^{(0)}}{d\varphi^2} + (1-m^2 B^2) u^{(0)} - \frac{M}{\mathcal{L^{\prime}}^2} = 0. \label{zz2}
\end{equation}
The solution of Eq.~(\ref{zz2}) is
\begin{equation}
u^{(0)} = \frac{m}{\mathcal{L^{\prime}}^2} \left[ 1 + e \, \cos \left( \sqrt{1-m^2 B^2}\, \phi \right) \right], \label{zz3}
\end{equation}
which is analogous to the Newtonian result. The integration constants are chosen as the orbital eccentricity $e$ (assumed small, similar to GR) and the initial condition $\phi_0 = 0$.
At first order in $\epsilon$, the differential equation reads
\begin{equation}
\frac{d^2u^{(1)}}{d\varphi^2} + (1-m^2 B^2) u^{(1)} - \frac{\mathcal{L'}^2}{m} (u^{(0)})^2 = 0, \label{zz4}
\end{equation}
which admits an approximate solution of the form
\begin{align}
u^{(1)} &\simeq \frac{m}{\mathcal{L^{\prime}}^2} \, e \, \phi \, \sqrt{1-m^2 B^2}\, \sin \left( \phi \, \sqrt{1-m^2 B^2}\right)\nonumber\\
&+ \frac{m}{\mathcal{L^{\prime}}^2} \left[ \frac{1+e^2}{2} - \frac{e^2}{6} \cos \left( 2 \phi \, \sqrt{1-m^2 B^2}\right) \right]. \label{zz5}
\end{align}

For practical purposes, the second term in Eq.~(\ref{zz5}) can be neglected, as it consists of a constant shift and an oscillatory contribution that averages to zero.
Therefore, the perturbative solution (\ref{kk4}) reads
\begin{align}
  u &\simeq \frac{m}{\mathcal{L^{\prime}}^2}\Bigg[1+ e \cos \left( \sqrt{1-m^2 B^2} \phi \right)\nonumber\\
  &+\epsilon e \phi \sqrt{1-m^2 B^2} \sin \left( \sqrt{1-m^2 B^2} \phi \right) \Bigg].\label{zz6}
\end{align}
Because \(\epsilon \ll 1\), the perturbative solution (\ref{zz6}) can be rewritten in the form of an ellipse equation,
\begin{equation}
  u \simeq \frac{m}{\mathcal{L^{\prime}}^2} \left[1+ e \cos \left(\sqrt{1-m^2 B^2} (1-\epsilon)\phi \right)\right].\label{zz7} 
\end{equation}

Due to the presence of the magnetic field, the orbit doesn't remain periodic but more than $2\pi$ for conical defect as shown in Eq.~(\ref{aa5}). Thus, the period $\Phi$ is given by  
\begin{equation}
\Phi = \frac{2 \pi (1+m^2 B^2)}{\sqrt{1-m^2 B^2} (1-\epsilon)} \approx 2 \pi + \Delta \Phi, \label{zz8}
\end{equation}
where $\Delta \Phi$ represents the advance of the perihelion.  By expanding to first order in the small parameters $m B$, and $\epsilon$, the perihelion shift $\Delta \Phi$ is given by
\begin{equation}
\Delta \Phi = 2 \pi \epsilon + 3 \pi m^2 B^2 = \Delta \Phi_{\rm GR} + \delta \Phi_{\rm B}, \label{zz9}
\end{equation}
where each term corresponds to a distinct contribution:  

(i) The standard general relativity contribution:
\begin{equation}
\Delta \Phi_{\rm GR} = 2 \pi \epsilon=\frac{6\pi G_N\,m}{c^2\,(1-e^2)\,a},\label{zz10}
\end{equation}
with $c$ being the speed of light, \(m\) the geometrical mass, \(e\) the orbital eccentricity and \(a\) being the semi-major axis of the orbital ellipse.

(ii) The contribution due to the magnetic field:
\begin{equation}
\delta \Phi_{\rm B} = 3 \pi m^2 B^2, \label{zz11}
\end{equation}
Equation~(\ref{zz9}) clearly shows that the uniform magnetic field introduce additional corrections to the standard general relativistic perihelion advance.

\section{Harmonic Oscillations}

To properly analyze the oscillations of test particles around circular orbits under perturbations, we employ the modified equations of motion near stable circular orbits. It should be noted that epicyclic motion, i.e., linear harmonic oscillations, occurs when a test particle is slightly displaced from its equilibrium position along a stable circular orbit in the equatorial plane.

In the case of harmonic oscillatory motion, the corresponding frequencies, as measured by a local observer, are given by \cite{Faizuddin2026,Ahmed2026}
\begin{align}
\omega_r^2 &= - \frac{1}{2} \frac{\partial^2 U_{\rm eff}(r,\theta)}{\partial r^2}\Bigg{|}_{r=r_c,\theta=\theta_0}, \label{radial}\\
\omega_\theta^2 &= \frac{1}{2 r^2 g_{rr}(r,\theta)} \frac{\partial^2 U_{\rm eff}(r,\theta)}{\partial \theta^2}\Bigg{|}_{r=r_c,\theta=\theta_0}, \label{latitudinal} \\
\omega_\phi^2 &= \left(\frac{d\phi}{d\tau}\right)^2\Bigg{|}_{r=r_c,\theta=\theta_0}, \label{orbital}
\end{align}
where $\omega_r$, $\omega_\theta$, and $\omega_\phi$ denote the radial, latitudinal, and orbital frequencies, respectively. These expressions allow one to determine the characteristic epicyclic frequencies of a test particle moving around a stable circular orbit of fixed radii $r=r_c$ in the background of a non-rotating selected black hole. 

The effective potential (\ref{bb6}) for non-equatorial plane ($\theta=\theta_0$, constant) can be expressed as, 
\begin{align}
U_{\rm eff}(r, \theta)=\frac{1+(B^2-\alpha^2) r^2}{\Omega^2(r, \theta)}\left[1+\frac{\Omega^2(r, \theta)}{r^2 P(\theta) \sin^2 \theta}\,\mathcal{L}^2\right]\,f(r)\label{bb6a}
\end{align}

Using this effective potential, one can determine the epicyclic frequencies associated with test particles. Noted that in the limit $\alpha=0$ and $\beta=0$, we find following expressions for the epicyclic frequencies:
\begin{align}
    \omega_r^2&=-\frac{f f''-2 f'^2+3 f f'/r}{2 f-r f'}=-\frac{m (r-6 m)}{r^3 (r-3 m)},\label{radial3}\\
    \omega_\theta^2 &=\frac{f'}{r (2 f-r f')}=\frac{m}{r^2 (r-3 m)},\label{latitudinal3}\\
    \omega_\varphi^2 &=\frac{f'}{2 r}=\frac{m}{r^3}\label{orbital3}
\end{align}
which are similar to the standard Schwarzschild black hole solution case.

\begin{figure*}[t]
    \centering
    \includegraphics[width=\textwidth]{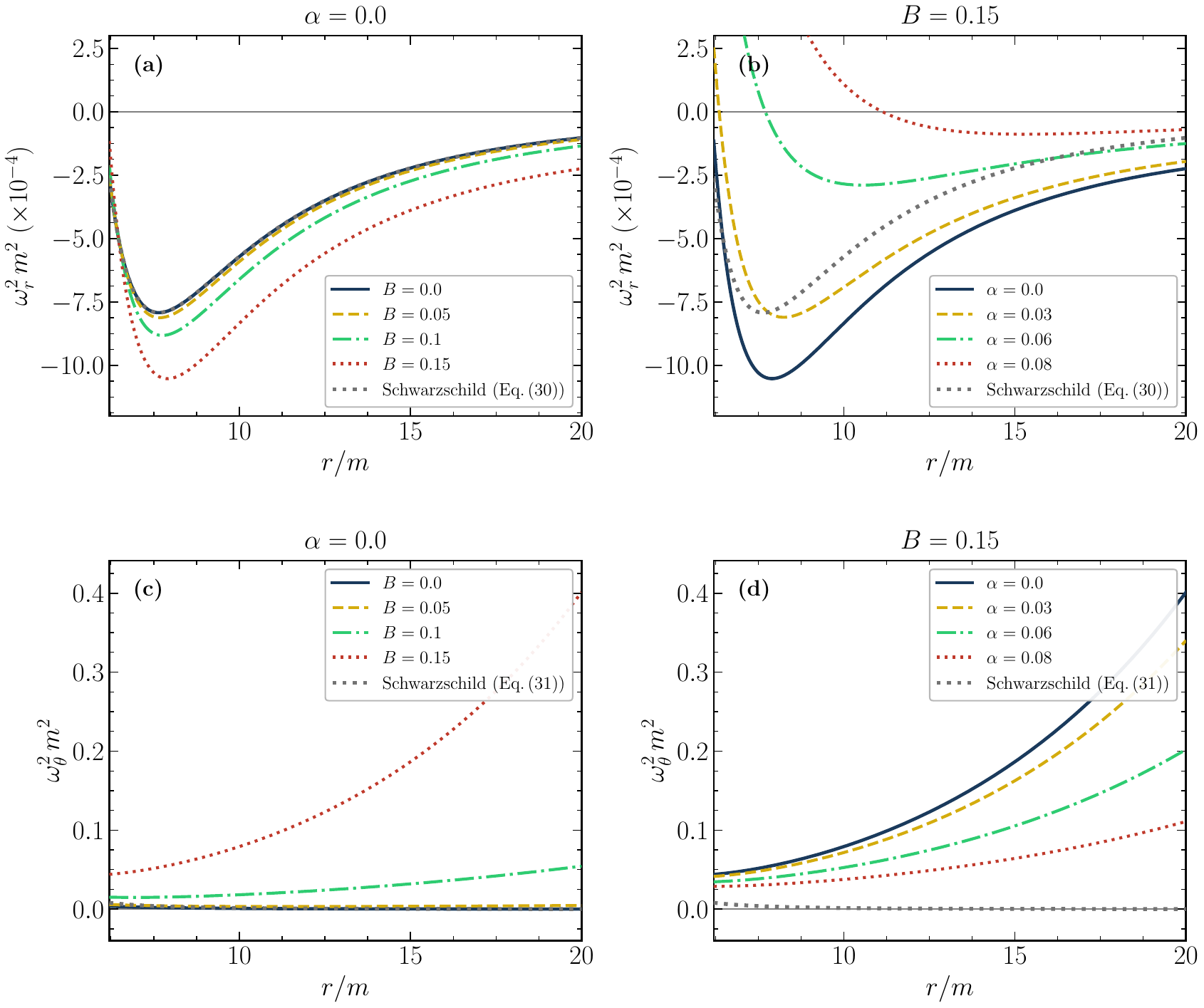}
    \caption{Radial and latitudinal epicyclic frequencies for small harmonic oscillations about equatorial circular orbits in the accelerating BR spacetime. The colored curves show the numerical evaluation of the definitions (\ref{radial}) and (\ref{latitudinal}), computed from the non-equatorial effective potential $U_{\rm eff}(r,\theta)$ in Eq.~(\ref{bb6a}) by taking second derivatives at $(r,\theta)=(r_c,\pi/2)$ along the family of circular orbits (i.e., with $\mathcal{L}=\mathcal{L}_{\rm sp}(r_c)$). Panels (a) and (b) display the radial frequency $\omega_r^2 m^2$ from Eq.~(\ref{radial}), while panels (c) and (d) display the latitudinal frequency $\omega_\theta^2 m^2$ from Eq.~(\ref{latitudinal}). In the left column, we vary the magnetic field $B\in\{0,\,0.05,\,0.10,\,0.15\}$ at fixed $\alpha=0$, whereas in the right column we vary the acceleration parameter $\alpha\in\{0,\,0.03,\,0.06,\,0.08\}$ at fixed $B=0.15$ (with $m=1$). The gray dotted curve provides the Schwarzschild benchmark: Eq.~(\ref{radial3}) in panels (a,b) and Eq.~(\ref{latitudinal3}) in panels (c,d).}
    \label{fig:epicyclic}
\end{figure*}

As shown in Fig.~\ref{fig:epicyclic}, the radial epicyclic frequency $\omega_r^2$ is negative in the stable-orbit domain, consistently with the sign convention in Eq.~(\ref{radial}): for a stable circular orbit the effective potential has a local minimum, $\partial_r^2U_{\rm eff}>0$, and therefore $\omega_r^2<0$. The latitudinal frequency $\omega_\theta^2$, on the other hand, measures the restoring response to small out-of-plane displacements and is expected to be positive whenever the equatorial plane is vertically stable, in accordance with Eq.~(\ref{latitudinal}). In all panels, the Schwarzschild limit is validated by the perfect overlap between the $(B,\alpha)=(0,0)$ curves and the dotted reference lines, namely Eq.~(\ref{radial3}) for the radial sector and Eq.~(\ref{latitudinal3}) for the latitudinal sector.

The four panels highlight how $B$ and $\alpha$ affect radial and vertical stability in distinct ways. In Fig. \ref{fig:epicyclic}(a) ($\alpha=0$ and varying $B$), increasing the magnetic field shifts the entire $\omega_r^2$ curve downward (more negative values) over most of the displayed range. Physically, this reflects a stronger curvature of the equatorial effective potential with respect to $r$ along the family of circular orbits: the BR magnetic deformation enhances the radial ``stiffness'' of small oscillations around the equilibrium radius. Close to the inner edge of the plot (near the onset of stable circular motion), the curves approach $\omega_r^2\to 0^-$, indicating marginal radial stability. This behavior is the epicyclic signature of the ISCO condition, since $\omega_r^2=0$ corresponds to the transition between stable and unstable circular orbits. At larger radii, the magnitude $|\omega_r^2|$ decreases and the curves tend back toward zero, consistent with the weakening of relativistic corrections and the progressive flattening of $U_{\rm eff}$.

Figure \ref{fig:epicyclic}(b) fixes $B=0.15$ and varies $\alpha$. Here, the acceleration parameter produces a qualitatively different trend: increasing $\alpha$ lifts the radial curve upward, making $\omega_r^2$ less negative and, for sufficiently large $\alpha$, pushing the curve closer to the marginal line $\omega_r^2=0$. In terms of the effective potential, $\alpha$ reduces the radial curvature at the minimum, softening the restoring force and therefore decreasing the radial epicyclic ``stiffness''. This is consistent with the role of acceleration in lowering and broadening the barrier in $U_{\rm eff}$ discussed earlier: a shallower minimum implies smaller $|\partial_r^2U_{\rm eff}|$ and hence a smaller magnitude of $\omega_r^2$. In particular, the progressive upward shift with $\alpha$ indicates that acceleration counteracts the magnetic confinement in the radial sector, moving the system closer to radial marginality over a wider range of radii.

The vertical sector is shown in Figs. \ref{fig:epicyclic}(c) and \ref{fig:epicyclic}(d). In Fig. \ref{fig:epicyclic}(c) ($\alpha=0$, varying $B$), the latitudinal frequency $\omega_\theta^2$ is positive and increases with radius, and its magnitude grows substantially as $B$ increases. This indicates that the magnetic deformation strengthens the effective restoring tendency against out-of-plane displacements: the second derivative $\partial_\theta^2U_{\rm eff}$ at $\theta=\pi/2$ increases in magnitude when the magnetic field is enhanced. Since the equatorial plane is a symmetry plane, the first derivative in $\theta$ vanishes there, and the sign and size of $\omega_\theta^2$ are controlled directly by the local curvature of $U_{\rm eff}$ in the $\theta$ direction. The monotonic growth with $r$ in these panels signals that, for the chosen parameter domain, vertical stability becomes increasingly rigid at larger radii, where the BR deformation contributes more strongly through the $\Omega$ and $P$ factors entering $U_{\rm eff}(r,\theta)$.

Figure \ref{fig:epicyclic}(d) fixes $B=0.15$ and varies $\alpha$, revealing the competing effect of acceleration on vertical stability. As $\alpha$ increases, the curves are systematically lowered: the vertical epicyclic frequency decreases, meaning that out-of-plane oscillations become softer. This trend is again understood in terms of the angular curvature of the effective potential: increasing $\alpha$ reduces the curvature of $U_{\rm eff}(r,\theta)$ with respect to $\theta$ at the equatorial plane, diminishing the restoring force. Importantly, the persistence of $\omega_\theta^2>0$ throughout the plotted range indicates that the equatorial plane remains vertically stable for these parameter choices, even though the restoring strength can be significantly reduced by acceleration. In this sense, Figs. \ref{fig:epicyclic}(b) and \ref{fig:epicyclic}(d) jointly show that $\alpha$ acts as a ``softening'' deformation in both radial and latitudinal sectors, whereas $B$ tends to increase the stiffness of the epicyclic motion.

Overall, Fig.~\ref{fig:epicyclic} provides a compact stability diagnostic: the radial frequency $\omega_r^2$ encodes the proximity to radial marginal stability (ISCO) via $\omega_r^2\rightarrow 0$, while the latitudinal frequency $\omega_\theta^2$ quantifies vertical confinement about the equatorial plane. The opposite trends induced by $B$ and $\alpha$ in each sector make clear that the external BR magnetic field and the acceleration parameter compete in shaping the local curvature of $U_{\rm eff}(r,\theta)$ and, consequently, the characteristic epicyclic response of massive circular orbits.

\section{Optical Properties} \label{sec4}

The optical appearance of a black hole, its shadow, and the deflection of light rays in its vicinity, constitute one of the most direct observational probes of the strong-field regime of gravity. In this section, we study null geodesics in the equatorial plane of the accelerating BR spacetime. We derive the effective potential for photons, obtain the photon sphere radius and shadow radius in closed form, analyze the photon trajectories for different values of the impact parameter, compute the radial force experienced by photons, and evaluate the Lyapunov exponent that governs the (in)stability of circular null orbits. Throughout, we examine how $B$ and $\alpha$ modify each of these quantities with respect to the Schwarzschild baseline.

\subsection{Effective potential for null geodesics}

We consider null geodesic motion in the equatorial plane defined by $\theta=\pi/2$ and $\dot{\theta}=0$. Setting $ds^2=\frac{1}{\Omega^2(r)}\,\tilde{ds}^2=0$ into the metric (\ref{aa1}) yields
\begin{equation}
-\frac{Q}{r^2}\,\dot{t}^2+\frac{r^2}{Q}\dot{r}^2+r^2\,\dot{\phi}^2=0.\label{dd1}
\end{equation}

Since the spacetime is independent of $t$ and $\phi$, we have two conserved quantities, the photon energy $\mathrm{E}$ and angular momentum $\mathrm{L}$:
\begin{equation}
\dot{t}=\frac{r^2\,\mathrm{E}}{Q},\qquad \dot{\phi}=\frac{\mathrm{L}}{r^2}.\label{dd2}
\end{equation}

Substituting into Eq.~(\ref{dd1}), the geodesic equation for the radial coordinate becomes
\begin{equation}
\dot{r}^2+\frac{Q\mathrm{L}^2}{r^4}= \mathrm{E}^2.\label{dd3}
\end{equation}

This equation can be re-written in the standard energy-conservation form:
\begin{equation}  
\dot{r}^2+V^\text{null}_\text{eff}(r)=\mathrm{E}^2,\label{dd4}
\end{equation}
where the null effective potential is
\begin{align}
    V^\text{null}_\text{eff}(r)&=\mathrm{L}^2\left[1-B^2 m^2 -\frac{2m}{r}  \right]\left[\frac{1}{r^2}+B^2-\alpha^2\right].\label{dd5}
\end{align}
In the Schwarzschild limit ($B=\alpha=0$), this reduces to $V^\text{null}_\text{eff}(r)=\mathrm{L}^2(1-2m/r)/r^2$, as expected. The magnetic field $B$ enters both through the overall prefactor $(1-B^2m^2)$ and through the term $(B^2-\alpha^2)r^2$, while $\alpha$ only enters through the latter combination, suggesting that the two parameters play complementary roles in modifying the potential.

\begin{figure*}[t]
    \centering
    \includegraphics[width=\textwidth]{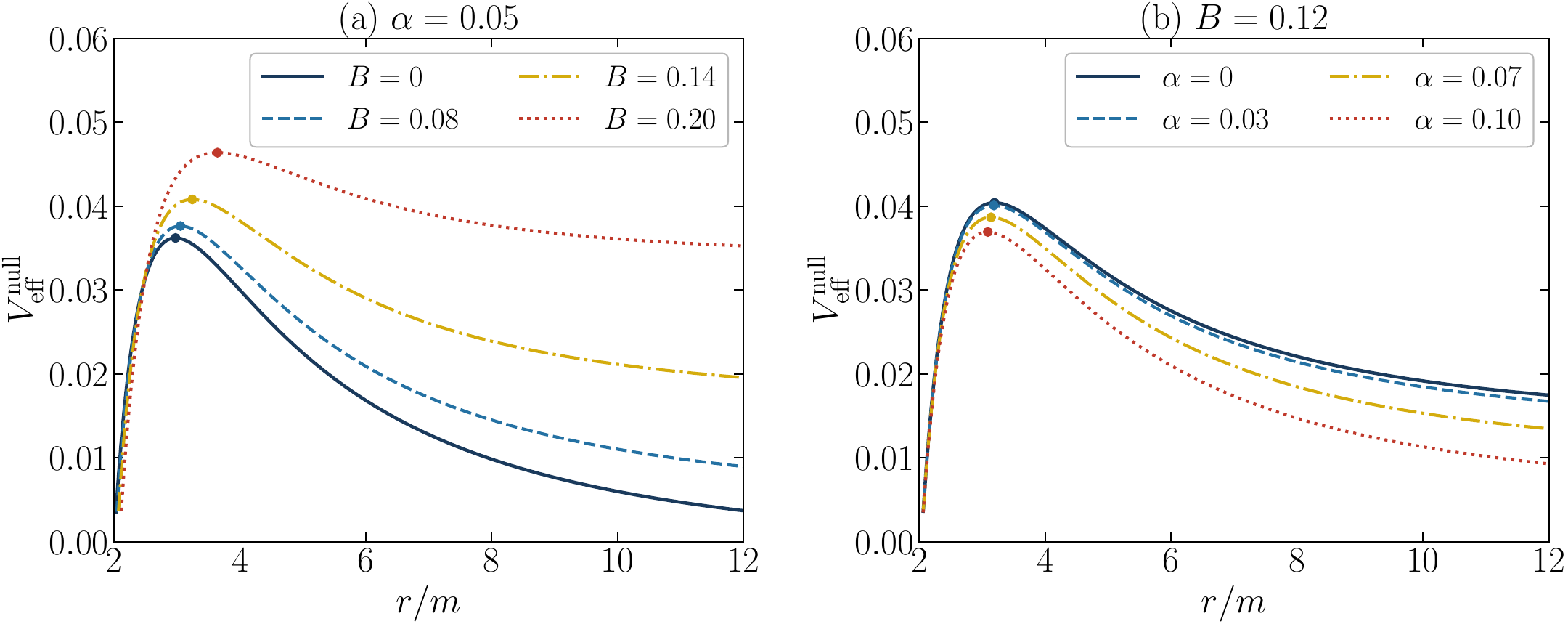}
    \caption{Null effective potential $V^{\rm null}_{\rm eff}(r)$ [Eq.~(\ref{dd5})] as a function of the radial coordinate $r/m$, with $\mathrm{L}=1$. Panel~(a) shows the variation with magnetic field strength $B \in \{0, 0.08, 0.14, 0.20\}$ at fixed $\alpha=0.05$; panel~(b) shows the variation with acceleration $\alpha \in \{0, 0.03, 0.07, 0.10\}$ at fixed $B=0.12$. Filled circles mark the local maxima, which coincide with the photon sphere radii $r_{\rm ph}$ at each parameter set. In panel~(a), increasing $B$ raises the potential barrier and shifts $r_{\rm ph}$ outward, indicating that the magnetic field confines the photon sphere to larger radii; the tail of the potential at large $r$ also increases, reflecting the growing importance of the $(B^2-\alpha^2)r^2$ term. In panel~(b), increasing $\alpha$ lowers the maximum and shifts $r_{\rm ph}$ inward, showing that acceleration defocuses photon trajectories and shrinks the photon sphere. The complementary nature of $B$ and $\alpha$ visible here translates directly into the opposing trends observed for the shadow radius in Fig.~\ref{fig:5}.}
    \label{fig:4}
\end{figure*}

The null effective potential is illustrated in Fig.~\ref{fig:4} for the same parameter ranges used in Fig.~\ref{fig:1}. The filled circles identify the photon sphere $r_{\rm ph}$ at each parameter set as the maximum of the potential. Figure \ref{fig:4}(a) shows that a larger magnetic field raises the potential barrier significantly and shifts $r_{\rm ph}$ to larger radii, while Fig. \ref{fig:4}(b) demonstrates the opposing effect of acceleration. The tails of the potential at large $r$ are also visibly affected: because $V^\text{null}_\text{eff} \propto (1+(B^2-\alpha^2)r^2)$, the effective potential grows asymptotically when $B^2 > \alpha^2$, which is the case for most curves shown.

\subsection{Circular null orbits: photon sphere}

Here, we study the circular null orbits and derive the key quantities associated with them. For circular photon orbits of radius $r=r_c$, we require $\dot{r}=0$ and $\ddot{r}=0$, which translate into the standard conditions on the effective potential:
\begin{equation}
    V^\text{null}_\text{eff}(r_c)=\mathrm{E}^2,\qquad \left.V^{\text{null}\,\prime}_\text{eff}(r)\right|_{r=r_c}=0.\label{dd3cc}
\end{equation}

The first condition gives the critical impact parameter $\beta_c \equiv \mathrm{L}/\mathrm{E}$ at which photons orbit the black hole:
\begin{align}
    \frac{1}{\beta_c}&=\frac{\mathrm{E}}{\mathrm{L}}\notag\\&=\frac{1}{r}\sqrt{\left(1-B^2 m^2 -\frac{2m}{r}\right)\left(1+\left(B^2-\alpha^2\right) r^2\right)}\Big{|}_{r=r_c}.\label{dd3dd}
\end{align}
The impact parameter depends on $B$, $\alpha$, and $m$: increasing $B$ enlarges $\beta_c$ while increasing $\alpha$ reduces it, reflecting the competition between the magnetic confinement and the acceleration-driven defocusing of photon trajectories.

The second condition $V^{\text{null}\,\prime}_\text{eff}(r)=0$ gives the photon sphere radius $r_\text{ph}$ as the solution of
\begin{equation}
    m\,(B^2-\alpha^2)\,r^2-(1-B^2\,m^2)\,r+3m=0.\label{dd3ee}
\end{equation}
Solving the above quadratic equation yields the photon sphere radius $r_{\rm ph}$. We note that the condition $m^2 B^2 < 1$ must be satisfied to ensure the existence of the event horizon, as discussed earlier. Below, we discuss some cases and determine the radius of photon sphere.

\begin{itemize}
\item $B=0,\,\alpha\ne 0$ (C-metric, no magnetic field), the photon sphere radius is given by 
\begin{align} 
r_\text{ph}=\frac{-1 + \sqrt{1+12\alpha^2m^2}}{2\alpha^2 m}.\label{dd3ff1} 
\end{align}
For small $\alpha m \ll 1$ this gives $r_\text{ph} \approx 3m - 9\alpha^2 m^3 + \mathcal{O}(\alpha^4 m^5)$, in agreement with perturbative results in the C-metric literature.
\item $\alpha=0,\,B\ne 0$ (Schwarzschild-BR configuration):
\begin{equation}
r_\text{ph}=\frac{1-B^2m^2+\sqrt{1-14B^2m^2+B^4m^4}}{2mB^2}.\label{dd3ff3}
\end{equation}
Noted here that the photon sphere exist provided \(m B \geq 5\) unless otherwise.

\item $\alpha=0,\,B=0$ (Schwarzschild black hole):
\begin{equation}
r_\text{ph}=3m.\label{dd3ff2}
\end{equation}
\end{itemize}

\begin{figure*}[ht!]
	\centering
	\includegraphics[width=\textwidth]{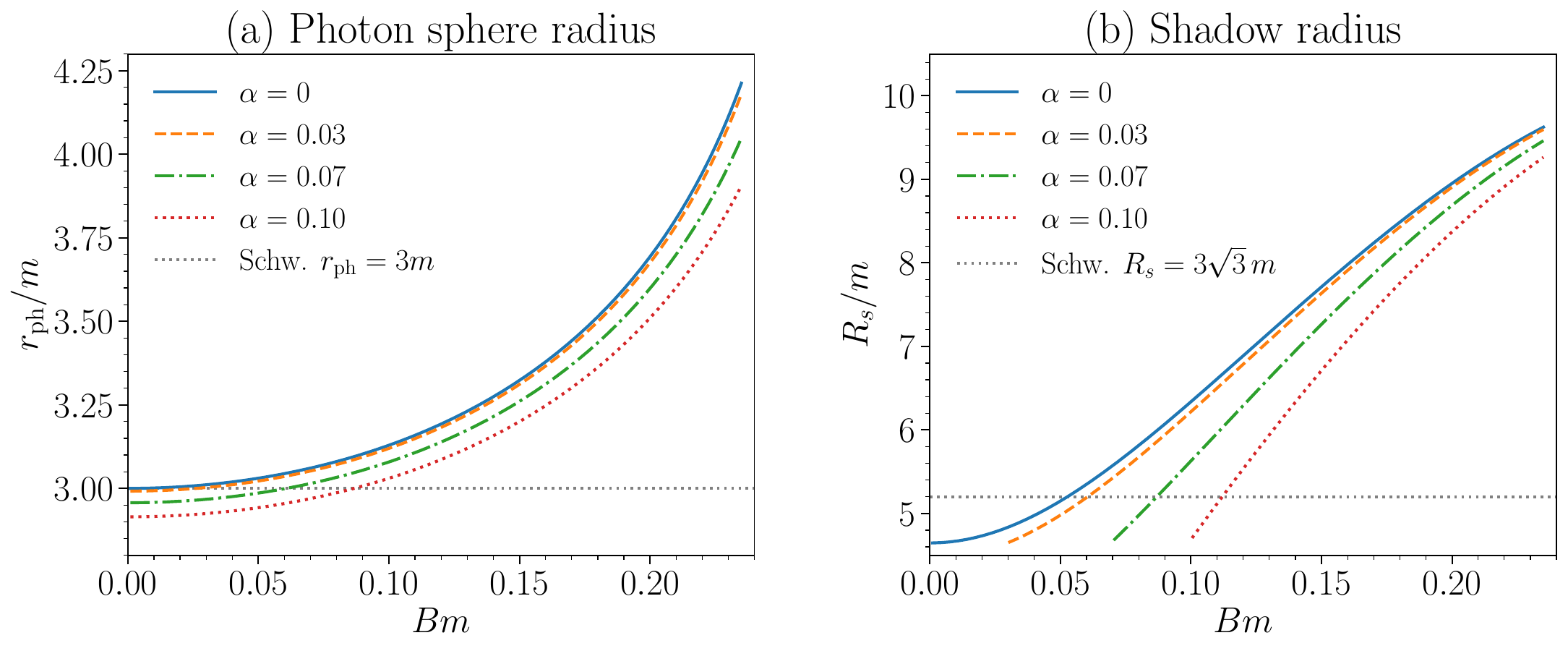}
	\caption{Photon sphere radius $r_{\rm ph}/m$ (panel~a, Eq.~(\ref{dd3ee})) and shadow radius $R_s/m$ (panel~b, Eq.~(\ref{shad}) with $r_O/m = 10$) as functions of the magnetic field parameter $Bm$, for $\alpha \in \{0,\,0.03,\,0.07,\,0.10\}$. The horizontal dotted lines indicate the Schwarzschild reference values $r_{\rm ph} = 3m$ and $R_s = 3\sqrt{3}\,m \approx 5.196\,m$. In panel~(a), $r_{\rm ph}$ increases monotonically with $B$ at fixed $\alpha$, growing from $3m$ to approximately $4.2m$ ($\alpha = 0$) or $3.9m$ ($\alpha = 0.10$) over $Bm \in (0, 0.235)$; at fixed $B$, increasing $\alpha$ reduces $r_{\rm ph}$, with the effect becoming more pronounced at larger $B$. In panel~(b), $R_s$ likewise grows monotonically with $B$ and decreases with $\alpha$, reaching values near $9.6m$ ($\alpha = 0$) and $9.3m$ ($\alpha = 0.10$) at $Bm = 0.235$; each curve is plotted only for $B \geq \alpha$, the range in which Eq.~(\ref{shad}) with $r_O/m = 10$ is physically well defined. The curves for different $\alpha$ values remain close together at small $B$ and diverge increasingly as $B$ grows, illustrating a non-trivial interplay between the two parameters. The combined reading of $r_{\rm ph}$ and $R_s$ could, in principle, disentangle the individual contributions of $B$ and $\alpha$ in EHT-type observations.}
	\label{fig:5}
\end{figure*}
The shadow radius for a static observer located at position $r_O$ is given by \cite{Perlick2022}
\begin{align}
R_{\rm sh}=r_{\rm ph}\sqrt{\frac{\left(1-B^2 m^2 -\frac{2m}{r_O}\right)\left(1+\left(B^2-\alpha^2\right) r^2_O\right)}{\left(1-B^2 m^2 -\frac{2m}{r_{\rm ph}}\right)\left(1+\left(B^2-\alpha^2\right) r^2_{\rm ph}\right)}}.\label{shad}
\end{align}

\begin{figure*}[tbhp]
	\centering
	\includegraphics[width=\textwidth]{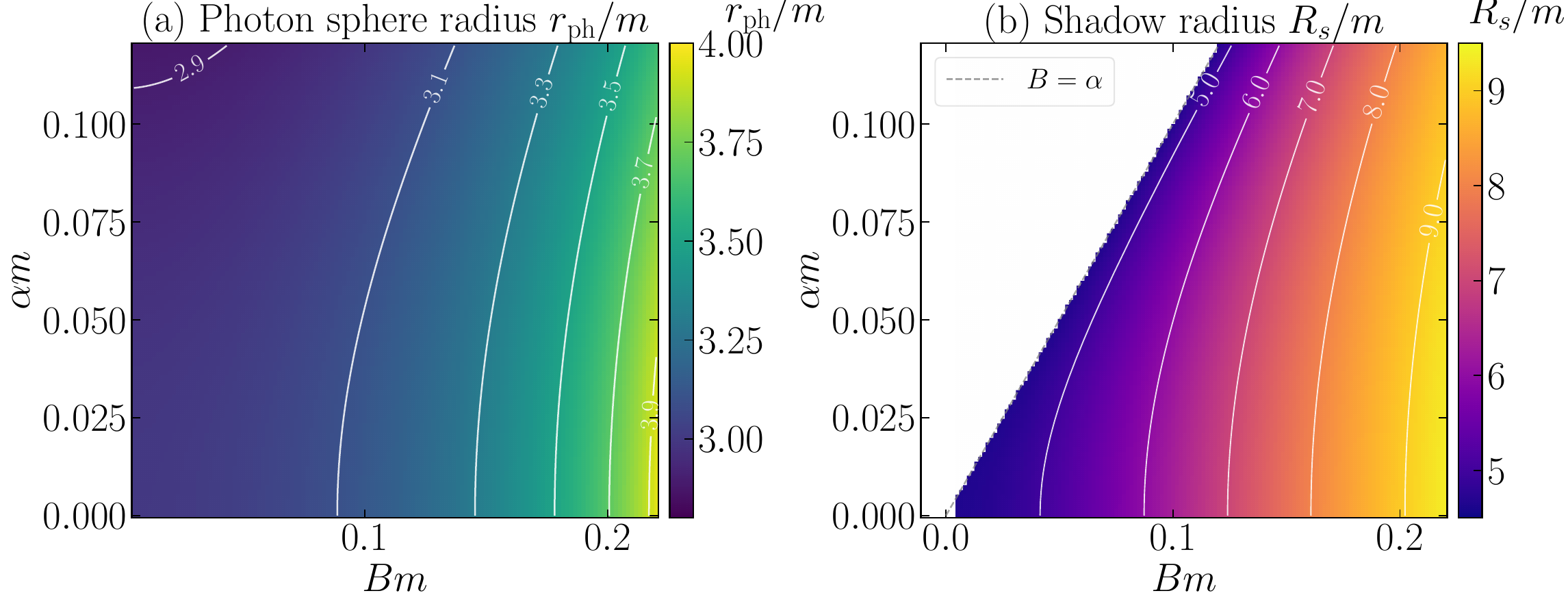}
	\caption{Heatmaps of the photon sphere radius $r_{\rm ph}/m$ (panel~a, Eq.~(\ref{dd3ee})) and shadow radius $R_s/m$ (panel~b, Eq.~(\ref{shad}) with $r_O/m = 10$) over the parameter plane $(Bm,\,\alpha m)\in[0.005,\,0.220]\times[0,\,0.120]$. White isocontours are drawn at the values indicated by the labels. In panel~(b) the lower-left triangular region $B < \alpha$ (separated by the dashed diagonal) is left blank because Eq.~(\ref{shad}) with $r_O/m = 10$ yields unphysical values there. Both observables increase with $B$ and decrease with $\alpha$; the magnetic field is the dominant driver ($\partial r_{\rm ph}/\partial(Bm) \approx 2.2$ times larger than $|\partial r_{\rm ph}/\partial(\alpha m)|$, and $\partial R_s/\partial(Bm) \approx 1.7$ times larger than $|\partial R_s/\partial(\alpha m)|$ at a representative interior point). The near-constancy of $R_s$ along the diagonal $B = \alpha$ ($4.65\,m \lesssim R_s \lesssim 4.71\,m$) contrasts with its rapid growth away from it, illustrating the non-trivial interplay between the two parameters.}
	\label{fig:9}
\end{figure*}
Numerical values of the photon sphere and shadow radius for representative values of the parameters are presented in Table~\ref{taba13}. One observes that both $r_\text{ph}$ and $R_\text{s}$ increase with $B$ at fixed $\alpha$, while they decrease with $\alpha$ at fixed $B$. This dual behavior is a distinctive feature of the accelerating BR geometry and distinguishes it from purely magnetic or purely accelerating deformations of the Schwarzschild metric.

\iffalse
\begin{table}[ht!]
\centering
\begin{tabular}{|c|cc|cc|cc|}
\hline
 & \multicolumn{2}{c|}{$B=0.1$} & \multicolumn{2}{c|}{$B=0.15$} & \multicolumn{2}{c|}{$B=0.2$} \\ 
\hline
$\alpha$ & $r_{\rm ph}$ & $R_{\rm s}$ & $r_{\rm ph}$ & $R_{\rm s}$ & $r_{\rm ph}$ & $R_{\rm s}$ \\ 
\hline
0.03 & 3.11977 & 5.03308 & 3.31135 & 5.26950 & 3.67510 & 5.54298 \\ 
0.05 & 3.10326 & 4.59996 & 3.29060 & 5.10043 & 3.64358 & 5.45055 \\ 
0.07 & 3.07915 & 3.83186 & 3.26046 & 4.82751 & 3.59844 & 5.30695 \\ 
0.09 & 3.04813 & 2.37104 & 3.22198 & 4.41989 & 3.54185 & 5.10467 \\ 
\hline
\end{tabular}
\caption{Numerical values of the photon sphere radius $r_{\rm ph}$ and shadow radius $R_{\rm s}$ for different values of $\alpha$ and $B$, with $m=1$.}
\label{taba13}
\end{table}
\fi

\begin{table*}[ht!]
\centering
\caption{Numerical values of photon sphere radius $r_{\rm ph}$ and shadow radius $R_{\rm sh}$ for various values of $\alpha$ at fixed $B$, with $r_O/m=10$.}
\label{taba13}
\begin{tabular}{|c|cc|cc|cc|cc|}
\hline
$mB$ & \multicolumn{2}{c|}{$m\alpha=0.03$} & \multicolumn{2}{c|}{$m\alpha=0.05$} & \multicolumn{2}{c|}{$m\alpha=0.07$} & \multicolumn{2}{c|}{$m\alpha=0.09$} \\
\cline{2-9}
 & $r_{\rm ph}/m$ & $R_{\rm sh}/m$ & $r_{\rm ph}/m$ & $R_{\rm sh}/m$ & $r_{\rm ph}/m$ & $R_{\rm sh}/m$ & $r_{\rm ph}/m$ & $R_{\rm sh}/m$ \\
\hline
0.1  & 3.11977 & 6.21809 & 3.10326 & 5.99478 & 3.07915 & 5.62911 & 3.04813 & 5.07036 \\
0.15 & 3.31135 & 7.63636 & 3.2906  & 7.49358 & 3.26046 & 7.26501 & 3.22198 & 6.92987 \\
0.2  & 3.6751  & 8.90636 & 3.64358 & 8.82229 & 3.59844 & 8.68865 & 3.54185 & 8.49526 \\
0.25 & 4.57571 & 9.82294 & 4.49058 & 9.7838  & 4.37716 & 9.72036 & 4.24627 & 9.62671 \\
\hline
\end{tabular}
\end{table*}

The behavior of $r_{\rm ph}$ and $R_s$ across the full range of $B$ and $\alpha$
is shown in Fig.~\ref{fig:5}.
Both quantities grow monotonically with $B$ at fixed $\alpha$
[Fig.~\ref{fig:5}(a)] and decrease with $\alpha$ at fixed $B$
[Fig.~\ref{fig:5}(b)], as anticipated from the analytical expressions.
For the photon sphere [Fig.~\ref{fig:5}(a)], $r_{\rm ph}$ grows from $3m$
to approximately $4.2m$ ($\alpha = 0$) or $3.9m$ ($\alpha = 0.10$) over the
range $Bm \in (0, 0.235)$, and the separation between curves of different
$\alpha$ widens with increasing $B$, signaling that the two parameters are not
simply additive in their effect on the photon sphere.
For the shadow radius [Fig.~\ref{fig:5}(b)], $R_s$ grows from
$\approx 3\sqrt{3}\,m$ to values approaching $9.6m$ ($\alpha = 0$) or
$9.3m$ ($\alpha = 0.10$) as $Bm$ increases to $0.235$; each curve is shown
only for $B \geq \alpha$, the regime in which Eq.~(\ref{shad}) with the
static observer at $r_O = 10m$ is physically well defined.
The Schwarzschild reference lines (dotted) confirm that both observables
recover the standard values at $B = \alpha = 0$, providing a consistency check.
These systematics suggest that simultaneous measurement of $r_{\rm ph}$ and
$R_s$ could, in principle, disentangle the individual contributions of $B$ and
$\alpha$.

Figure~\ref{fig:9} presents the joint parameter dependence of
$r_{\rm ph}$ and $R_s$ as continuous heatmaps over the two-dimensional
domain $(Bm,\,\alpha m) \in [0.005,\,0.220]\times[0,\,0.120]$.
This representation complements the line plots of Fig.~\ref{fig:5} by
making the interplay between the two parameters immediately visible across
the full parameter space.

In Fig. \ref{fig:9}(a) (photon sphere radius), the color encodes $r_{\rm ph}/m \in [2.88,\, 3.94]$ over the entire grid. The dominant gradient runs along the $B$-axis: at fixed $\alpha$, the photon sphere grows monotonically with $B$, from $r_{\rm ph} \approx 3.0\,m$ near $B = 0$ to $r_{\rm ph} \approx 3.94\,m$ at $(Bm,\,\alpha m) = (0.220,\, 0)$. The influence of $\alpha$ is secondary but systematic: increasing $\alpha$ at fixed $B$ shifts $r_{\rm ph}$ to smaller values, with the effect becoming progressively more pronounced at large $B$. At the upper-right corner of the domain, $(Bm,\,\alpha m) = (0.220,\,0.120)$, the photon sphere shrinks to $r_{\rm ph} \approx 3.62\,m$, compared with $3.94\,m$ at the same $B$ but $\alpha = 0$ — a reduction of roughly $8\%$ induced solely by the acceleration parameter. The white isocontours, spaced at intervals of $0.2\,m$, run nearly parallel to the $\alpha$-axis at small $B$ and tilt progressively toward the lower-right as $B$ increases, reflecting the growing non-additivity of the two parameters: a numerical sensitivity analysis at $(Bm,\,\alpha m) = (0.10,\, 0.06)$ gives $\partial r_{\rm ph}/\partial(Bm) \approx 2.67$ and $\partial r_{\rm ph}/\partial(\alpha m) \approx -1.21$, so the magnetic field
drives the photon sphere approximately twice as efficiently as the acceleration parameter suppresses it.

In Fig. \ref{fig:9}(b) (shadow radius), the lower-left triangular region $B < \alpha$, where the static-observer
formula~(\ref{shad}) with $r_O/m = 10$ yields unphysical values (see Sec.~\ref{sec:shadow}), is left blank. In the physically well-defined region $B \geq \alpha$, $R_s/m$ ranges from $\approx 4.65\,m$ along the diagonal $B = \alpha$ (marked by the dashed line) to $\approx 9.36\,m$ at the corner $(Bm,\,\alpha m) = (0.220,\, 0)$. The dominant gradient is again along the $B$-direction: at $\alpha = 0$, the shadow radius grows from $R_s \approx 4.65\,m$ at small $B$ to $9.36\,m$ at $Bm = 0.220$, crossing the reference value $3\sqrt{3}\,m \approx 5.196\,m$ near $Bm \approx 0.041$ and reaching $7\,m$ near $Bm \approx 0.124$.
The dependence on $\alpha$ at fixed $B$ is weaker but non-negligible: along the right edge $Bm = 0.220$, increasing $\alpha m$ from $0$ to $0.120$ reduces $R_s$ from $9.36\,m$ to $8.67\,m$, a suppression of roughly $7\%$. The isocontours of $R_s$ are closely spaced near the diagonal $B = \alpha$, where $R_s$ varies little (between $4.65\,m$ and $4.71\,m$ along the entire diagonal), and spread out rapidly as $B$ increases above $\alpha$, signaling
that $R_s$ is far more sensitive to $B$ than to $\alpha$ in this regime. A local sensitivity analysis at $(Bm,\,\alpha m) = (0.10,\, 0.06)$ yields $\partial R_s/\partial(Bm) \approx 31.2$ versus $\partial R_s/\partial(\alpha m) \approx -18.3$, a ratio of approximately $1.7$, consistent with the visual impression that the isocontours run at a
moderate angle to the $B$-axis.

Taken together, the two panels confirm that $B$ is the primary driver of both observables while $\alpha$ acts as a subdominant but measurable correction. The qualitatively different morphology of the two heatmaps, gradual and
nearly uniform in Fig. \ref{fig:9}(a), steep and concentrated near the $B$-axis in Fig. \ref{fig:9}(b), suggests that a joint measurement of $r_{\rm ph}$ and $R_s$ would provide complementary constraints on the two parameters, potentially allowing them to be disentangled in future EHT-type observations.

\subsection{Photon trajectories}

The motion of photons in the equatorial plane can be fully characterized by the ratio $b \equiv \mathrm{L}/\mathrm{E}$, the impact parameter. The orbit equation is obtained by eliminating the affine parameter $\lambda$ from Eqs.~(\ref{dd2}) and (\ref{dd3}), giving
\begin{align}
    \left(\frac{dr}{d\varphi}\right)^2 &= \frac{r^4}{b^2}\left(1 - B^2m^2 - \frac{2m}{r}\right)\left(1+(B^2-\alpha^2)r^2\right) \notag\\ &- r^4 V^\text{null}_\text{eff}(r) \cdot \frac{1}{\mathrm{L}^2}\cdot \frac{r^4}{\mathrm{E}^2},
\end{align}
which simplifies to the compact form
\begin{equation}
\left(\frac{du}{d\varphi}\right)^2 = \frac{1}{b^2} - u^2\left(1-B^2m^2-2mu\right)\left(1+\frac{B^2-\alpha^2}{u^2}\right),\label{orbit_eq}
\end{equation}
where $u = 1/r$ is the inverse radial coordinate. Three qualitatively distinct types of photon orbits exist, depending on the value of $b$ relative to the critical impact parameter $\beta_c$:

\begin{itemize}
\item \textit{Unstable circular orbit} ($b = \beta_c$): 

the photon asymptotically spirals toward the photon sphere located at $r = r_{\rm ph}$.

\item \textit{Scattering orbits} ($b > \beta_c$): 
photons coming from infinity reach a minimum radial distance $r_{\rm min} > r_{\rm ph}$ and are then deflected back to infinity. The deflection angle diverges as $b \to \beta_c^{+}$.

\item \textit{Capture orbits} ($b < \beta_c$): 
photons cross the event horizon and are captured by the black hole without returning to infinity.
\end{itemize}

The deflection angle for a photon with $b > \beta_c$ passing through the equatorial plane can be written as
\begin{align}
\hat{\Phi}(b) = 2\int_{r_{\rm min}}^{\infty} \frac{dr}{\sqrt{r^4/b^2 - r^2 f(r)\left(1+(B^2-\alpha^2)r^2\right)}} - \pi,\label{defl_angle}
\end{align}
where $f(r) = 1 - B^2m^2 - 2m/r$ and $r_{\rm min}$ is the largest root of the integrand denominator. In the weak-field limit $m/b \ll 1$, this integral can be expanded to give
\begin{equation}
\hat{\Phi} \approx \frac{4m}{b} + \left[\frac{15\pi m^2}{4b^2} + \pi(B^2-\alpha^2)\right] + \mathcal{O}\!\left(\frac{m^3}{b^3}\right),
\end{equation}
showing that the magnetic field increases the bending angle while acceleration reduces it, consistent with the behavior of $r_\text{ph}$ found above.

\begin{figure}[t!]
\centering
\includegraphics[width=\columnwidth]{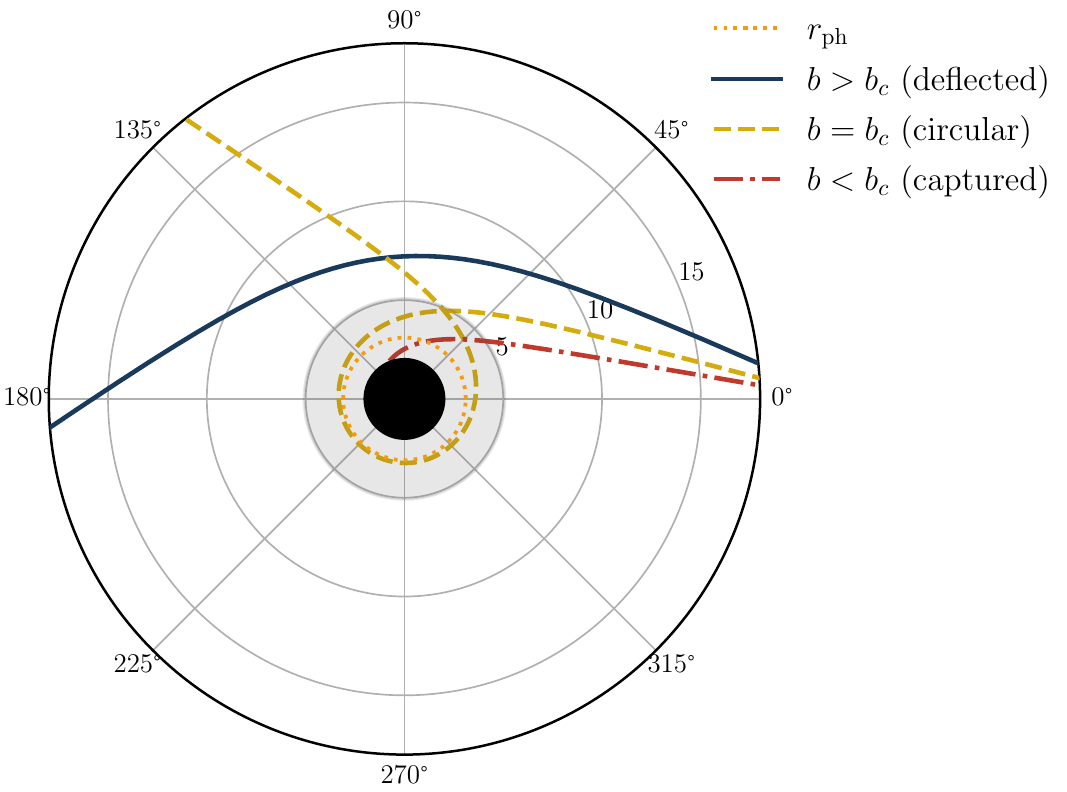}
\caption{Photon trajectories in the equatorial plane of the accelerating BR spacetime with $B=0.10$ and $\alpha=0.05$, displayed in polar coordinates $(r,\varphi)$. The filled black disk represents the black hole interior ($r < r_h$). The orange dotted circle marks the photon sphere at $r_{\rm ph} \approx 3.10\,m$. The shaded gray region inside the dashed circle represents the black hole shadow of radius $R_s$, i.e.\ the angular region from which no photon escapes to a distant observer. Three representative trajectories are shown, integrated numerically from $r_{\rm start} = 22m$ inward: the solid dark curve ($b > b_c$) is a deflected (scattering) orbit that reaches a minimum approach radius and returns to infinity after being significantly bent; the dashed amber curve ($b = b_c$) is the unstable circular orbit that asymptotically approaches $r_{\rm ph}$ after completing approximately one full revolution; the dash-dotted red curve ($b < b_c$) is a capture orbit that plunges through the photon sphere and falls into the black hole. The critical impact parameter is $b_c = r_{\rm ph}/\sqrt{f(r_{\rm ph})(1+(B^2-\alpha^2)r_{\rm ph}^2)} \approx 5.19\,m$ for these parameter values.}
\label{fig:6}
\end{figure}

The three classes of photon trajectories are displayed in Fig.~\ref{fig:6} for $B=0.10$, $\alpha=0.05$. The deflected orbit (dark solid curve) passes to the right of the black hole, reaches a turning point, and returns to infinity with a net bending angle. The nearly-circular orbit (amber dashed) spirals inward but barely escapes, winding almost once around the photon sphere before returning, illustrating the logarithmic divergence of the deflection angle as $b \to b_c^+$. The captured orbit (red dash-dotted) crosses $r_{\rm ph}$ and ends at the horizon; it has negligible angular extent because its impact parameter lies well below the critical value.

\begin{figure*}[t!]
\centering
\includegraphics[width=\textwidth]{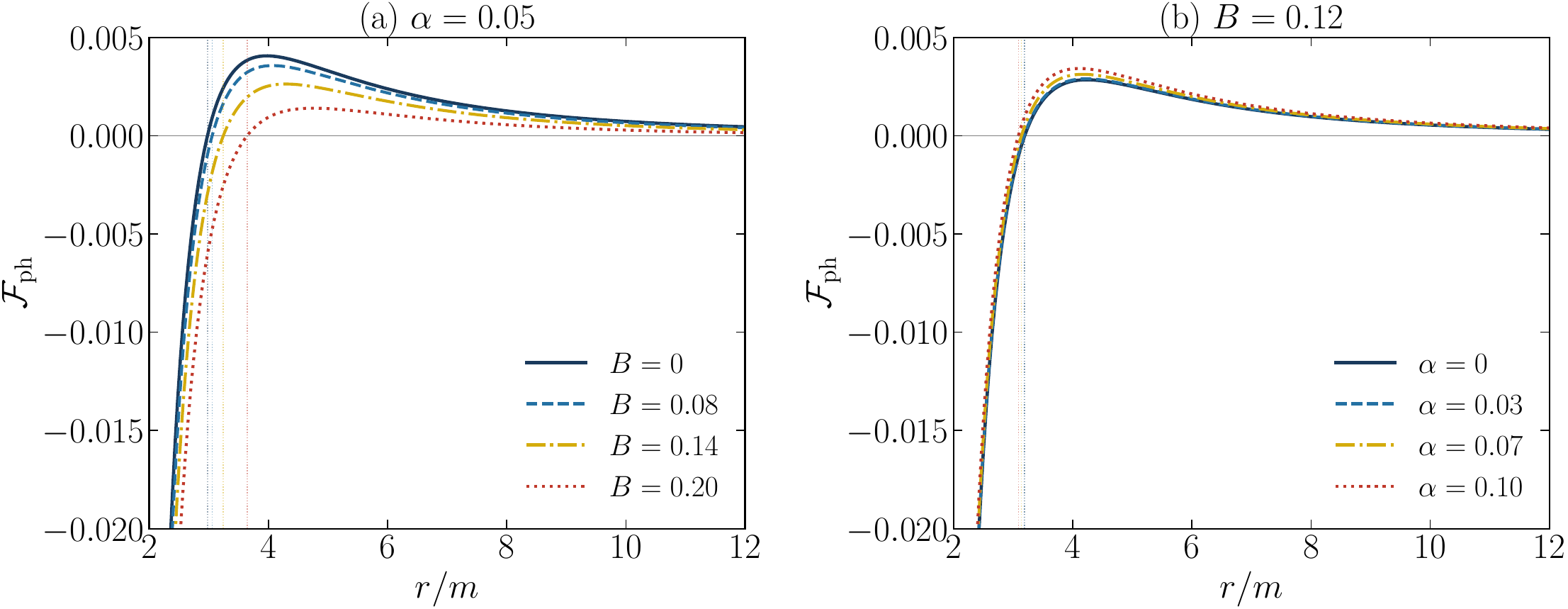}
\caption{Effective radial force on photons $\mathcal{F}_{\rm ph}$ [Eq.~(\ref{dd6}), with $\mathrm{L}=1$] as a function of $r/m$. Panel~(a) shows results for four values of the magnetic field $B \in \{0, 0.08, 0.14, 0.20\}$ at fixed $\alpha=0.05$; panel~(b) shows results for four values of the acceleration $\alpha \in \{0, 0.03, 0.07, 0.10\}$ at fixed $B=0.12$. Dotted vertical lines indicate the respective photon sphere radii $r_{\rm ph}$ for each curve, which coincide exactly with the zeros of $\mathcal{F}_{\rm ph}$: the force is negative (attractive) for $r < r_{\rm ph}$ and positive (repulsive) for $r > r_{\rm ph}$. In panel~(a), a larger $B$ shifts the zero to a larger $r$, in accordance with the growth of $r_{\rm ph}$ with $B$ shown in Fig.~\ref{fig:5}(a). In panel~(b), increasing $\alpha$ shifts the zero inward and slightly reduces the depth of the attractive well. The maximum of the repulsive part near $r \approx 4$--$5\,m$ reflects the restoring force that binds the circular photon orbit; its slight reduction with increasing $\alpha$ is consistent with the lowering of the potential barrier seen in Fig.~\ref{fig:4}(b).}
\label{fig:7}
\end{figure*}

\subsection{Effective Radial Force Experienced by Photons}

Having characterized the photon trajectories, it is instructive to examine the effective force acting on photons in the gravitational and magnetic field of the accelerating BR black hole. This force quantifies the tendency of the effective potential to attract or repel photons and is defined as
\begin{equation}
    \mathrm{F}_\text{ph}=-\frac{1}{2}\,\frac{\partial V^\text{null}_\text{eff}}{\partial r}.\label{force}
\end{equation}
Using the expression (\ref{dd5}) for the null effective potential, we find
\begin{eqnarray}
\mathrm{F}_\text{ph}=\frac{\mathrm{L}^2}{r^4}\,\left(r-m\left( 3+B^2r(m+r)-\alpha^2r^2 \right)\right).\label{dd6}
\end{eqnarray}

The sign of $\mathrm{F}_\text{ph}$ determines whether the photon is attracted toward ($\mathrm{F}_\text{ph} < 0$) or repelled from ($\mathrm{F}_\text{ph} > 0$) the black hole. The force vanishes at the photon sphere $r = r_\text{ph}$, as expected. For $r > r_\text{ph}$ the force is repulsive, maintaining the circular orbit against inward drift, while for $r < r_\text{ph}$ it is attractive, driving photons toward the singularity. One can verify that increasing $B$ displaces the zero of $\mathrm{F}_\text{ph}$ outward, consistent with the enlargement of $r_\text{ph}$, while increasing $\alpha$ shifts it inward.

The radial profiles of $\mathcal{F}_{\rm ph}$ are shown in Fig.~\ref{fig:7}. The change of sign at $r_{\rm ph}$ (marked by dotted vertical lines) is clearly visible in both panels: the force is strongly attractive near the horizon and transitions to repulsive just outside the photon sphere. The position of the zero moves outward with increasing $B$ [Fig. \ref{fig:7}(a)] and inward with increasing $\alpha$ [Fig. \ref{fig:7}(b)], consistent with the trends in $r_{\rm ph}$ established in the previous subsection.

\begin{figure}[ht!]
\centering    
\includegraphics[width=\columnwidth]{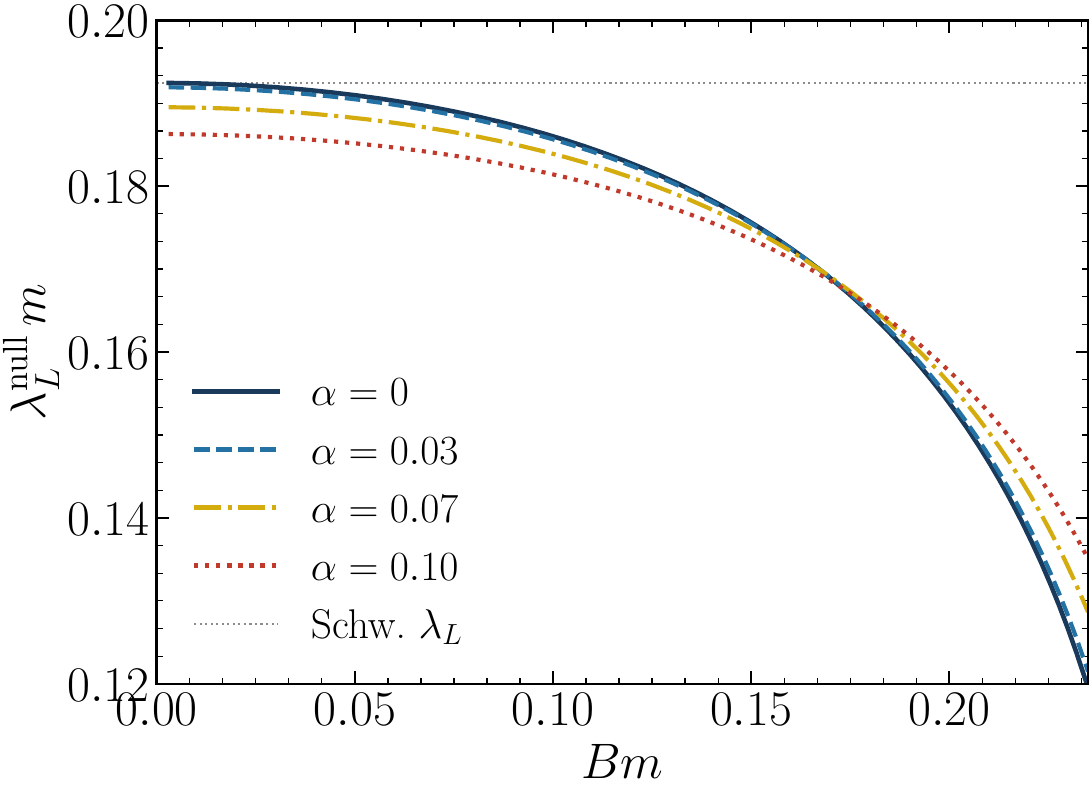}
\caption{Lyapunov exponent $\lambda_L^{\rm null}\,m$ [Eq.~(\ref{cond2})] governing the instability of the photon sphere, plotted as a function of the magnetic field parameter $Bm$ for $\alpha \in \{0, 0.03, 0.07, 0.10\}$. The horizontal dotted line marks the Schwarzschild reference value $\lambda_L^{\rm Sch} = 1/(3\sqrt{3}\,m) \approx 0.1925/m$. All curves start near the Schwarzschild value at $Bm=0$ and decrease monotonically with increasing $B$: the magnetic field enlarges the photon sphere, thereby increasing the orbital radius and reducing the curvature of the effective potential at its maximum, which in turn lowers the instability rate. Increasing $\alpha$ at fixed $B$ lifts the curves slightly (faster destabilization), consistent with the contraction of $r_{\rm ph}$ with $\alpha$. The eikonal correspondence $\omega_I \approx -\lambda_L$ connects these results directly to the imaginary part of the quasinormal mode spectrum, so Fig.~\ref{fig:8} effectively maps how the gravitational-wave ringdown decay rate is modified by the external magnetic field and acceleration.}
\label{fig:8}
\end{figure}

\subsection{Lyapunov Exponent}

We now turn our attention to the stability of the photon sphere. The stability of circular null orbits can be quantified through the Lyapunov exponent $\lambda_L$, which characterizes the exponential rate at which nearby trajectories diverge from the circular orbit. A real and positive $\lambda_L$ indicates orbital instability (i.e., the photon sphere is unstable, as is generically the case for black hole photon spheres), while an imaginary $\lambda_L$ would signal a stable circular orbit.

For circular null orbits, the Lyapunov exponent is expressed in terms of the second derivative of the effective potential as \cite{Cardoso2009}
\begin{equation}
\lambda^\text{null}_L=\sqrt{-\frac{V^{\text{null}\,\prime\prime}_\text{eff}(r_\text{ph})}{2\,\dot{t}^2}},\label{cond1}
\end{equation}
where $\dot{t}$ is given in Eq.~(\ref{dd2}) evaluated at $r = r_\text{ph}$.

Computing $V^{\text{null}\,\prime\prime}_\text{eff}$ from (\ref{dd5}) and using $\dot{t} = r_\text{ph}^2(1+B^2r_\text{ph}^2)\mathrm{E}/Q(r_\text{ph})$, we find after simplification:
\begin{align}
\lambda^\text{null}_L = \frac{1}{r_\text{ph}^2}\sqrt{\frac{Q(r_\text{ph})}{1+B^2r_\text{ph}^2}}\,\sqrt{\mathcal{N}(r_\text{ph})},\label{cond2}
\end{align}
where the stability function $\mathcal{N}(r)$ is defined by
\begin{align}
\mathcal{N}(r) &= \frac{2m}{r}\left(3 + B^2r(m+r) - \alpha^2 r^2\right) - 1 \nonumber\\
&\quad - r^2\left(B^2-\alpha^2\right)\left(1-B^2m^2-\frac{2m}{r}\right).\label{calN}
\end{align}
One can check that $\mathcal{N}(3m) = 1 > 0$ in the Schwarzschild limit ($B=\alpha=0$), giving $\lambda_L^\text{Sch} = (1/3m)\sqrt{1/3m} = 1/(3m\sqrt{3m})$, consistent with the known result. Since $\mathcal{N}(r_\text{ph}) > 0$ for all parameter values considered, the photon sphere is always unstable, which is the physically expected result for a black hole spacetime.

The Lyapunov exponent has a direct physical interpretation: its inverse $\tau_{\rm instab} = 1/\lambda_L$ sets the timescale on which small perturbations grow, and it is related to the imaginary part of the quasinormal mode frequencies in the eikonal limit \cite{Cardoso2009} via $\omega_I = -\lambda_L$. Both $B$ and $\alpha$ modify $\lambda_L$: increasing $B$ tends to decrease the exponent (more stable photon sphere, relatively speaking), while increasing $\alpha$ increases it (faster destabilization).

The behavior of the Lyapunov exponent across the full parameter range is shown in Fig.~\ref{fig:8}. The exponent decreases monotonically with $B$, indicating that the photon sphere becomes relatively more stable (a longer instability timescale) as the magnetic field increases. The sensitivity to $\alpha$ is mild at small $B$ but becomes discernible at larger $B$, where the four curves separate visibly. The recovery of the Schwarzschild value at $B=0$ for all $\alpha$ values confirms the correct limiting behavior.

To provide a global view of how the photon-sphere and shadow observables depend on the two parameters simultaneously, Fig.~\ref{fig:9} presents two-dimensional color maps of $r_{\rm ph}(B,\alpha)$ and $R_s(B,\alpha)$ over the physically relevant parameter space. The labeled iso-contours allow one to read off, at a glance, which combinations of $B$ and $\alpha$ yield a given value of either observable. The near-diagonal orientation of the contours implies that $B$ and $\alpha$ produce partially compensating effects on both $r_{\rm ph}$ and $R_s$, so a single observation of the shadow size alone cannot uniquely fix both parameters without additional information.

\section{Rate of Energy Emission}

Quantum effects in curved spacetime reveal that black holes are not perfectly black; instead, they emit a thermal flux from the region near the event horizon, with a temperature determined by the surface gravity. In semiclassical terms, this process can be understood as a tunneling or particle-production phenomenon near the horizon, resulting in a gradual decrease of the black hole’s mass as energy is radiated to infinity \cite{Javed2019}. For a distant observer, the observed radiation depends not only on the Hawking temperature but also on the likelihood that the emitted particles overcome the black hole’s gravitational potential.

In the high-frequency (geometric-optics) limit, the absorption cross section exhibits oscillations around a constant value, denoted $\sigma_{\rm lim}$. Since the capture of energetic quanta is largely determined by null geodesics, the black hole’s shadow directly influences the high-energy absorption cross section. The constant limiting value of the cross section, which is connected to the radius of the photon sphere, is given by \cite{Misner1973,Mashhoon1973,Wei2013}:
\begin{equation}
\sigma_{\rm lim}\approx \pi r^2_{\rm ph},\label{ee1}
\end{equation}
where \(r_{\rm ph}\) denotes the radius of the photon sphere.

Within this approximation, the spectral energy emission rate of black hole is given by the following equation \cite{Wei2013}
\begin{equation}
\frac{d^{2}\mathbb{E}}{d\omega\,dt}=\frac{2\pi ^{2}\sigma _{\rm lim}}{e^{\omega/T}-1}\,\omega ^{3},\label{ee2}
\end{equation}
where $\omega$ is the emitted frequency and $T$ is the Hawking temperature.

Substituting $T$ form Eq.~(\ref{zz5}), we find the rate of spectral energy as,
\begin{equation}
\frac{d^{2}\mathbb{E}}{d\omega\,dt}=\frac{2\pi ^{3} r^2_{\rm ph}\,\omega ^{3}}{\left[\exp\left\{\frac{8\pi m \omega}{(1-m^2 B^2)^2+4 m^2 (B^2-\alpha^2)}\right\}-1\right]},\label{ee3}
\end{equation}
where $r_{\rm ph}$ can be determined using Eq.~(\ref{dd3ee}).

\begin{figure}[ht!]
    \centering
    \includegraphics[width=0.95\linewidth]{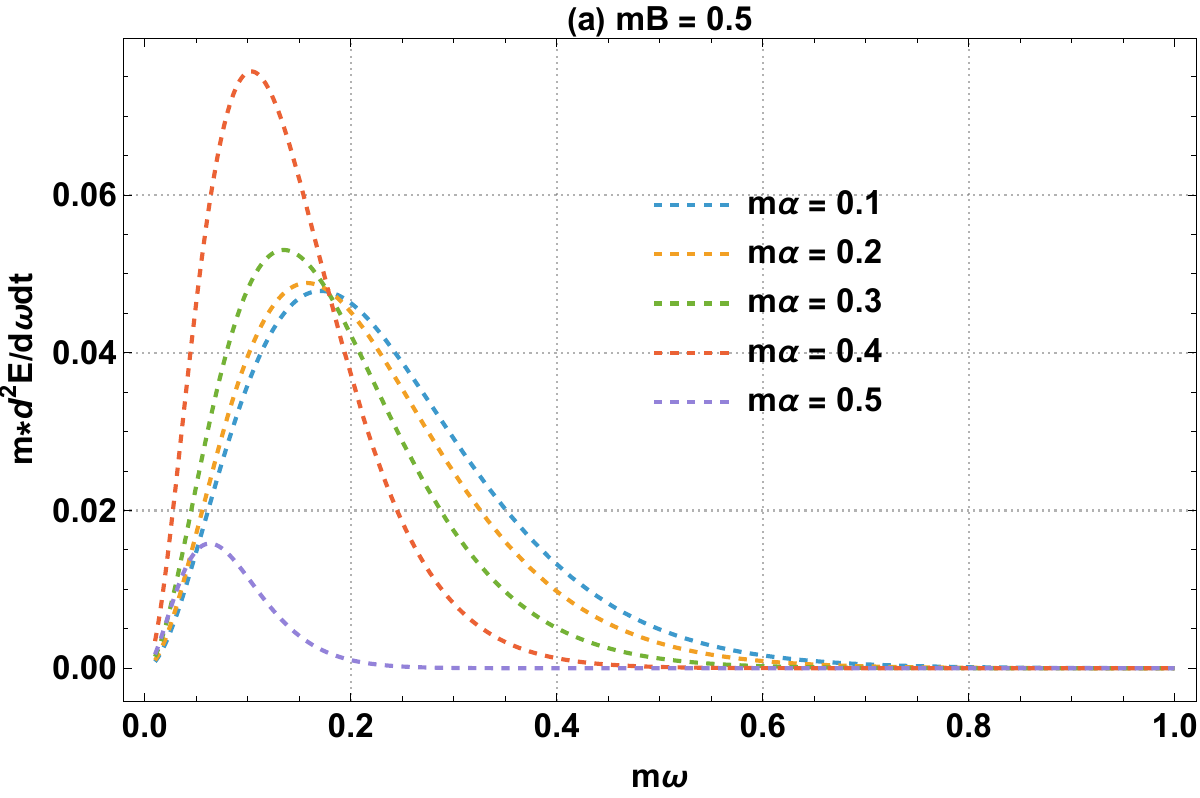}
    \includegraphics[width=0.95\linewidth]{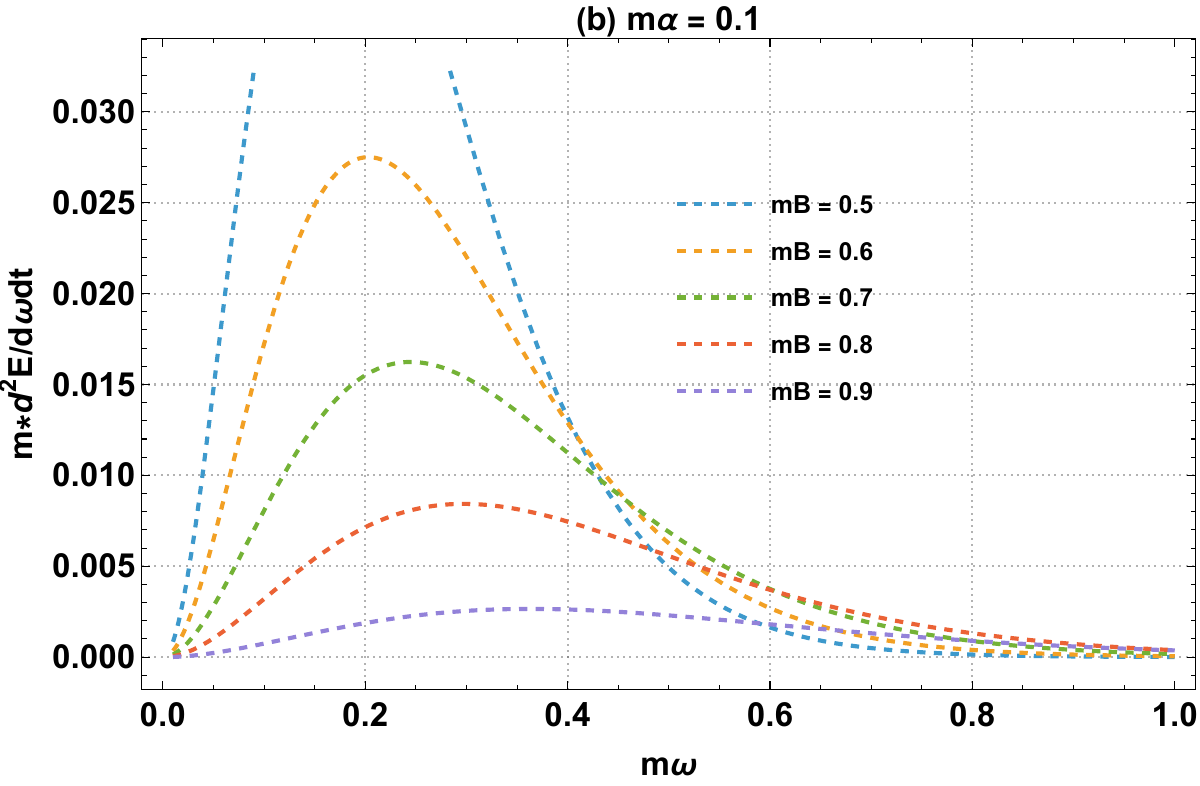}  
    \caption{The behavior of the energy emission rate as a function of frequency $\omega$ by varying $\alpha$ and $B$.}
    \label{fig:energy-emission}
\end{figure}

In Fig.~\ref{fig:energy-emission}, we illustrate the variation of the energy emission rate as a function of the dimensionless frequency $m\omega$ for different values of the magnetic field $B$ and the acceleration parameter $\alpha$.

Top panel shows that the energy emission rate initially increases with increasing $m\omega$, reaches a maximum at a certain frequency, and then gradually decreases for larger values of $m\omega$. The height of this peak increases with increasing $\alpha$ up to $\alpha = 0.4/m$. However, for $\alpha > 0.4/m$, the peak value decreases sharply. This indicates that moderate acceleration enhances the emission rate, whereas larger acceleration suppresses it significantly.

In contrast, bottom panel exhibits the same qualitative behavior with respect to $m\omega$: the emission rate rises to a maximum and then decreases. However, in this case, increasing the magnetic field $B$ leads to a monotonic reduction in the peak height. Thus, although the overall spectral profile remains unchanged, the magnetic field suppresses the maximum energy emission rate, in contrast to the non-monotonic influence of the acceleration parameter of small value.

\section{Conclusions}\label{summary}

In this work, we presented a systematic study of the accelerating BR black hole spacetime \cite{Ovcharenk2026}, focusing on how the external magnetic field strength $B$ and the acceleration parameter $\alpha$ jointly modify orbital dynamics, epicyclic stability, optical observables, and thermodynamic behavior with respect to the Schwarzschild baseline. This geometry interpolates between several well-known limits, including the Schwarzschild, C-metric, and Schwarzschild--BR configurations, and therefore provides a unified framework to disentangle magnetic and acceleration effects in a single exact vacuum solution.

In the timelike sector, we derived the equatorial effective potential and obtained closed expressions for the specific angular momentum $\mathcal{L}_{\rm sp}$ and specific energy $\mathcal{E}_{\rm sp}$ of circular orbits [Eqs.~(\ref{bb8}) and (\ref{bb10})], which were used to identify the marginal stability condition defining the ISCO. Our numerical scan shows a clear competition between the two deformations: increasing $B$ generally pushes the ISCO to larger radii, reflecting the enhanced effective confinement, whereas increasing $\alpha$ tends to shift the ISCO inward and partially offsets the magnetic contribution (Fig.~\ref{fig:3}). We further illustrated representative massive-particle trajectories by integrating the orbit equation, highlighting how $(B,\alpha)$ reshape turning points and the separatrix between scattering and bound motion (Fig.~\ref{fig:orbits_eq19}).

A key extension of the present version is the epicyclic analysis. Using the non-equatorial effective potential $U_{\rm eff}(r,\theta)$ [Eq.~(\ref{bb6a})], we computed the radial and latitudinal epicyclic frequencies associated with small harmonic oscillations about equatorial circular orbits [Eqs.~(\ref{radial}) and (\ref{latitudinal})]. The resulting profiles (Fig.~\ref{fig:epicyclic}) provide a compact stability diagnostic: the approach of $\omega_r^2$ to zero signals the onset of radial marginal stability, while $\omega_\theta^2$ quantifies the restoring strength against out-of-plane perturbations. We find that the magnetic field typically increases the stiffness of the epicyclic response, whereas acceleration softens both radial and vertical oscillations, consistent with the way $(B,\alpha)$ reshapes the curvature of $U_{\rm eff}(r,\theta)$.

In the null sector, we derived the effective potential for photons [Eq.~(\ref{dd5})] and obtained analytic expressions for the photon sphere radius and the corresponding shadow radius for a static observer [Eqs.~(\ref{dd3ee}) and (\ref{shad})]. The numerical results confirm that both $r_{\rm ph}$ and $R_s$ increase with $B$ and decrease with $\alpha$ (Table~\ref{taba13}, Fig.~\ref{fig:5}). We also classified photon trajectories (Fig.~\ref{fig:6}), computed the effective radial force on photons [Eq.~(\ref{dd6}) and Fig.~\ref{fig:7}], and evaluated the Lyapunov exponent controlling the instability of circular null orbits [Eq.~(\ref{cond2}) and Fig.~\ref{fig:8}]. To provide a global view suitable for parameter inference, we presented two-dimensional parameter-space maps for $r_{\rm ph}(B,\alpha)$ and $R_s(B,\alpha)$ (Fig.~\ref{fig:9}), which reveal partially compensating trends between $B$ and $\alpha$.

Finally, we analyzed the thermodynamic sector by computing the horizon radius from $Q(r_h)=0$ [Eq.~(\ref{zz1})] and the Hawking temperature [Eq.~(\ref{zz5})]. Although the horizon location depends on $B$ but not on $\alpha$, the temperature depends on both parameters through the surface gravity, showing that acceleration affects the black hole thermal behavior even when it does not move the horizon (Fig.~\ref{fig:temeprature}). 

Overall, the accelerating BR spacetime exhibits a rich and highly parameter-sensitive structure. The external magnetic field and acceleration play distinct and often competing roles: $B$ tends to enhance confinement, enlarge the photon sphere and shadow, and strengthen epicyclic stiffness, while $\alpha$ typically produces the opposite trends by softening orbital stability and shrinking optical observables. These signatures suggest that combined constraints from ISCO-related observables, epicyclic stability, and black hole shadow measurements could, in principle, help disentangle the individual contributions of $B$ and $\alpha$ in this class of accelerating, magnetically deformed black hole spacetimes.

Several extensions are natural. One may generalize the present analysis to the charged sector, investigate quasinormal modes beyond the eikonal approximation, and explore additional lensing observables (e.g.\ relativistic Einstein rings) that are directly tied to the strong-field photon dynamics and the instability scale set by $\lambda_L$.

\section*{Acknowledgments}

A.A.-B. acknowledges the support of Al-Hussein Bin Talal University. F.A. acknowledges the Inter University Center for Astronomy and Astrophysics (IUCAA), Pune, India for selecting visitors associtaeship. E. O. Silva acknowledges the support from grants CNPq/306308/2022-3, FAPEMA/UNIVERSAL-06395/22, and (CAPES) - Brazil (Code 001).

\end{document}